\documentclass[a4paper,11pt]{article}
\usepackage{amsfonts,amsmath,amssymb,bm,epsfig,url}
\usepackage[dvips]{color}

\usepackage{dcolumn}
\usepackage[pagebackref=false, colorlinks=true]{hyperref}
\usepackage{textcomp}
\usepackage{float}
\usepackage{subfigure}
\usepackage{graphicx}
\usepackage[normalem]{ulem}  
\usepackage[square,sort,comma,numbers]{natbib}
\definecolor{redish}{rgb}{0.7,0.2,0.0}  
\definecolor{bluish}{rgb}{0.2,0.5,0.8}

\hypersetup{linkcolor=redish,          
                  citecolor=blue,        
                  filecolor=magenta,      
                  urlcolor=bluish}

\def \o{\omega}

\def \O{\Omega}

\def \p{\partial}
\def \d{\Delta}

\def \th{\theta}

\def \f{\frac}

\title{Circular orbits in Kerr-Taub-NUT spacetime and their implications for
accreting black holes and naked singularities}
 \author{Chandrachur Chakraborty\footnote{chandra@pku.edu.cn}
\\{\it Kavli Institute for Astronomy and Astrophysics, Peking University,
Beijing 100871, China}
\\
{}
\\
Sudip Bhattacharyya
\\{\it Department of Astronomy and Astrophysics,} \\{\it Tata Institute
of Fundamental Research, Mumbai 400005, India}}

\date{}

\begin{document}

\maketitle

\begin{abstract}
It has recently been proposed that the accreting collapsed object
GRO J1655--40 could contain a non-zero gravitomagnetic monopole, and 
hence could be better described with the more general Kerr-Taub-NUT 
(KTN) spacetime, instead of the Kerr spacetime. This makes the KTN 
spacetime astrophysically relevant. In this paper, we study properties 
of various circular orbits in the KTN spacetime, and find the 
locations of circular photon orbits (CPOs) and 
innermost-stable-circular-orbits (ISCOs). Such orbits are important to 
interpret the observed X-ray spectral and timing properties of accreting 
collapsed objects, viz., black holes and naked singularities. Here we 
show that the usual methods to find the ISCO radius do not work for 
certain cases in the KTN spacetime, and we propose alternate ways.
For example, the ISCO equation does not give any positive real radius solution
for particular combinations of Kerr and NUT parameter values for
KTN naked singularities. In such a case, accretion efficiency generally 
reaches $100\%$ at a particular orbit of radius $r=r_0$, and hence we choose $r=r_0$ 
as the `ISCO' for practical purposes.
 
\end{abstract}

\section{Introduction}

The primary purpose of this paper is to study the stable circular orbits (SCOs) in the 
Kerr-Taub-NUT (KTN) spacetime, which is a geometrically stationary and axisymmetric 
vacuum solution of the Einstein equation. The marginally stable circular orbit (also 
known as innermost-stable-circular-orbit or ISCO) plays a very important role in the 
relativistic astrophysics and for accreting X-ray 
sources \cite{jh}. It is well-known that equatorial circular orbits with $r \geq r_{\rm ISCO}$ ($r_{\rm ISCO}$ is
the radius of ISCO) are stable, whereas those with $r < r_{\rm ISCO}$ are unstable. Accretion flows of 
the free matter can continue its circular motion in the orbits $r \geq r_{\rm ISCO}$ and it 
faces radial free-fall for $r < r_{\rm ISCO}$. However, to study the SCOs 
in the KTN spacetime one should first start with the study of geodesics. The preliminary studies
on geodesics in the KTN spacetime was done in \cite{cc2, abj} but there was no detailed discussion on
the SCOs. The aim of this paper is to find the SCOs which are physically realistic 
and also relevant for the accretion flows of the free matter. These orbits could be
extremely important as they are astrophysically relevant for studying the accretion 
physics as well as some other astrophysical phenomena in KTN spacetime. We show that all SCOs,
which are derived from the usual stability analyses 
of the orbits, cannot exist in the KTN spacetime. Rather, some SCOs are not feasible, 
and the innermost SCO (or, we can call it as `physical ISCO') is determined from the 
remaining SCOs. The unfeasibility happens due to the special 
geometric structure of the KTN spacetime, which we discuss in sections \ref{fr}--\ref{sec:mb}. 
We show that the test particle (or, we can say accreting matter) moving in the KTN 
spacetime cannot continue its stable circular motion at the unfeasible SCOs.
Mainly, the unusual behaviours in the ergoregion, Kepler frequency and the energy 
of the orbits prevent the test particle to have all of the SCOs. The same situation does
not arise in the Kerr spacetime, as well as some other known stationary
and axisymmetric spacetimes. Here, one immediate question is, 
which fundamental entity is responsible for this peculiarity? The answer is the 
gravitomagnetic monopole or the so-called NUT parameter.

Historically, Newmann, Unti and Tamburino (NUT) \cite{nut} discovered
a stationary and spherically symmetric \cite{mis, lnbl} vacuum solution (which is now known as the NUT solution)
of the Einstein's equation, that contains the gravitomagnetic monopole
or NUT parameter. This solution is related to neither merely post-Newtonian
nor some modified theory \cite{rs,rs2}. However, Demianski and Newman found that the
NUT spacetime is in fact produced by a `dual mass' \cite{dn} or the 
gravitomagnetic charge/monopole. NUT spacetime is the generalized version of the Schwarzschild spacetime
with the non-zero NUT charge and if the NUT charge vanishes, the NUT solution reduces to
the well-known Schwarzschild solution. If the Kerr spacetime contains the NUT parameter
or vice-versa, it is regarded as the KTN spacetime.
Gravitomagnetic monopole is basically the gravitational analogue
of Dirac's magnetic monopole \cite{dirac, saha} and Bonnor \cite{bon} physically interpreted it 
as `a linear source of pure angular momentum' \cite{dow,rs}, i.e., `a massless rotating rod'.
The gravitomagnetic monopole is a fundamental aspect of physics and the
Einstein-Hilbert action requires no modification \cite{rs2} to accommodate it.

Lynden-Bell and Nouri-Zonoz \cite{lnbl} were perhaps the first to motivate the investigation
on the observational possibilities for gravitomagnetic monopoles. They suggested
that the signatures of gravitomagnetic monopole might be found in the spectra of supernovae, quasars, 
or active galactic nuclei \cite{lnbl,kag}. They also studied the effects of 
gravitomagnetic monopole aka NUT charge on light rays as a gravitational lens and microlens 
\cite{nzl, nzml}. In another work, the local velocity of an orbiting star
in the equatorial plane of a Kerr and a KTN black hole was studied in relation to its spectral line
shifts, as measured by the distant observers \cite{nz1}. Interestingly, it was also proposed that 
the charged perfect fluid disks could be the sources of Taub-NUT-type spacetimes \cite{gg}, and
the KTN solution representing relativistic thin disks could be
of great astrophysical importance \cite{gg, liu, pra}. 
In fact, several interesting observational consequences were proposed \cite{liu, wei, zak,lcj}
for the KTN spacetime in last few years but the practical ways to detect it, were not 
proposed \cite{cbgm}.
 
KTN spacetime is the stationary and axisymmetric vacuum solution of the Einstein equation. 
As we know that the axisymmetric vacuum solutions of Einstein equation are used to describe 
a wide range of black holes appear in the Universe and the Pleba\'nski-Demia\'nski (PD) metric is 
the most general solution until now \cite{hl}. Schwarzschild, Kerr, Kerr-Newman, Taub-NUT, KTN, 
Reissner-Nordstr\"om, and all other well-known vacuum solutions are the special cases 
of this PD metric. Among all these solutions, the most prominent solution is Kerr spacetime, as it is 
astrophysically relevant. However, in a very recent paper \cite{cbgm}, the first
observational indication of the gravitomagnetic monopole has been reported and this makes
the KTN spacetime astrophysically relevant. Based on the X-ray observations of an 
astrophysical collapsed object, black hole (BH) or naked singularity (NS),
GRO J1655--40, it has been inferred there that this object contains the non-zero 
gravitomagnetic monopole. It was found earlier that the three independent 
primary X-ray observational methods gave significantly different spin values for the above
mentioned accreting collapsed object. Employing a new technique, Ref. \cite{cbgm} has demonstrated 
that the inclusion of one extra parameter (i.e., gravitomagnetic monopole or
NUT parameter $n$) not only makes the spin
and other parameter values inferred from the three methods 
consistent with each other, but also makes the inferred black hole mass consistent
with an independently measured value. Therefore, it may be advantageous to use the
more general KTN spacetime for accreting collapsed objects, such as GRO J1655--40.
This motivates us to study SCOs and CPOs in KTN spacetime.

The scheme of the paper is as follows. In section \ref{ktn}, we briefly recapitulate
the KTN spacetime. We discuss the usual methods to obtain the radii of SCOs and
the ISCO in section \ref{sec:isco}. These methods are based on the stability analyses of 
the circular orbits, which are derived from the effective potential formalism, as well
as from the radial epicyclic frequency. The detailed discussions on the locations of SCOs 
and/or the ISCO 
are covered in section \ref{sco}. As we have already mentioned, some SCOs
are unfeasible and we discuss this process in sections 
\ref{fr}--\ref{sec:mb}, which is extremely important for the KTN spacetime.
The similar studies of sections \ref{fr}--\ref{sec:mb} could also
be relevant in future to find the accessible SCOs in other 
spacetimes. Section \ref{cpo} is devoted to the radii of CPOs in 
the KTN BHs and KTN NSs. Finally, we conclude
in section \ref{dis}.

\section{{Brief description of Kerr-Taub-NUT spacetime}}\label{ktn}
The metric of the KTN spacetime can be expressed as \cite{ml, nz1}
\begin{eqnarray}
ds^2=-\f{\d}{p^2}(dt-A d\phi)^2+\f{p^2}{\d}dr^2+p^2 d\th^2
+\f{1}{p^2}\sin^2\th(adt-Bd\phi)^2
\label{metric}
\end{eqnarray}
with 
\begin{eqnarray}\nonumber
\d&=&r^2-2Mr+a^2-n^2, \,\,\,\,\,\,\,\,\,  p^2=r^2+(n+a\cos\th)^2,
\\
A&=&a \sin^2\th-2n\cos\th, \,\,\,\,\,\,\,\,  B=r^2+a^2+n^2
\end{eqnarray}
where $M$ is the mass, $a_*=a/M$ is the Kerr parameter 
and $n_*=n/M$ is the NUT parameter. Setting $\d=0$ and $g_{tt}=0$, one can obtain the 
radii of the outer horizon and outer ergoregion as
\begin{eqnarray}
 r_h = M(1+\sqrt{1+n_*^2-a_*^2}) \,\,\,\, 
 {\rm and,} \,\,\,\, r_{e} = M(1 + \sqrt{1+n_*^2-a_*^2 \cos^2\th})
 \label{ergo}
\end{eqnarray}
respectively. These two quantities are relevant for the astrophysical purposes. 
One can see that the radius 
of the ergoregion depends on the value of $n_*$ in the equatorial plane :
$r_e|_{\th=\pi/2}=M(1 + \sqrt{1+n_*^2})$, whereas it becomes $2M$ \cite{ckj} for all
values of $a_*$ in the Kerr spacetime. Interestingly, $p^2$ vanishes at \cite{mcd}
\begin{eqnarray}
r=0 \,\,\,\,\, {\rm and} \,\,\,\, \th = \cos^{-1}(-n_*/a_*),
\label{sing}
\end{eqnarray}
which indicates the location of singularity in KTN spacetime. Therefore, the above 
expression (eq. \ref{sing}) reveals that the singularity 
does not arise for $n_* > a_*$, which indicates a singularity-free KTN BH,
whereas for a KTN BH with $n_* = a_*$, singularity arises at $\th=\pi$, covered by the 
horizon. However, the singularity always arises for the $n_* < a_*$
case which could be a KTN BH or a KTN NS depending on the numerical values of $a_*$ and $n_*$.
As the horizon ($r_h$) vanishes for $a_* > \sqrt{1+n_*^2}$, one can always 
obtain a KTN NS in this case, whereas a KTN BH with singularity (covered by the horizon)
arises if and only if the following condition is satisfied : 
$n_* \leq a_* \leq \sqrt{1+n_*^2}$.

\section{Basic discussions for obtaining the innermost stable circular orbits}\label{sec:isco} 
It is well-known that a thin accretion disk can extend up to ISCO, and 
not inside ISCO. We note that, throughout this paper, the ISCO and other 
discussed limiting equatorial circular orbits are considered the
innermost edge of a thin accretion disk. In this section, we briefly describe
the two ways to derive the so-called ISCO equation, which are based on the usual stability 
analyses of the orbits. One way to derive it is from the expression of the radial
epicyclic frequency ($\O_r$), and another way is from the effective potential formalism.
However, as the expression of $\O_r$ can be derived from
the effective potential \cite{abkl}, these two approaches are not independent to each other.

\subsection{Effective potential formalism and stable circular orbits}\label{effp}
We will not repeat the whole derivation in this paper, but one 
can easily derive the ISCO equation for KTN spacetime from the expression of effective potential
($V_{\rm eff}$) which can be expressed as (see eq. (71) of \cite{cc2})
\begin{equation}
 V_{\rm eff}(r,E,L)=\f{1}{2}\left[(E^2-1)-\f{P^2-(r^2+n^2+O^2)\d}{(r^2+n^2)^2}\right]
\label{ep}
\end{equation}
where 
\begin{eqnarray}
 P(r)=BE-La \,\,\,\, {\rm and} \,\,\,\,  O(r)=L-aE.
\end{eqnarray}
$E$ and $L$ indicate the energy and angular momentum per unit mass
of a test particle which orbits around a KTN collapsed object.
For $n=0$, the effective potential reduces to eq. (15.20) of \cite{jh}
which is valid in the Kerr spacetime. One intriguing feature of $V_{\rm eff}$
is that it is finite at $r=0$ due to the presence of NUT charge, whereas it 
diverges in the cases of Schwarzschild and Kerr geometries.
This can be found from the following expression :
\begin{eqnarray}
 V_0=V_{\rm eff}|_{r=0}=\f{a^2(1-3E^2)+4aEL-L^2}{2n^2}-1.
 \label{v0}
\end{eqnarray}
One should note that $r=0$ is relevant only for the KTN NS case,
because $r=0$ is hidden inside the horizon in the BH case. However, 
if $n$ vanishes, $V_{\rm eff}$ will diverge at $r=0$, which could be realized
from the Kerr geometry, as $r=0, \th=\pi/2$ represents the ring singularity
\cite{ckj,ckp}. Such a singularity does not exist at
that particular point ($r=0, \th=\pi/2$) in the KTN geometry. This would be the main 
reason to obtain a finite effective potential at $r=0$. Another interesting point is 
that $V_0$ does not depend on the mass ($M$) explicitly, rather it
depends only on the values of $a_*$, $n_*$, $E$ and $L/M$. 

Now, for a particle to rotate in a circular orbit at radius $r=R$, its initial radial 
velocity ($\dot{r}$) must vanish. Imposing this condition in (see eq. (15.19) of \cite{jh})
\begin{equation}
 \f{E^2-1}{2}=\f{1}{2}\dot{r}^2+V_{\rm eff}(r,E,L)
\label{ep0}
\end{equation}
we obtain 
\begin{equation}
 \f{E^2-1}{2}=V_{\rm eff}(R,E,L).
\label{ep2}
\end{equation}
To stay in a circular orbit the radial acceleration must also vanish.
Thus, differentiating eq. (\ref{ep0}) with respect to $r$ leads to the 
condition:
\begin{equation}
 \f{\partial V_{\rm eff}(r,E,L)}{\partial r}|_{r=R}=0.
\label{ep3}
\end{equation}
Stable orbits are the ones for which small radial displacements away from
$R$ oscillate about it rather than accelerate away from it. Just as in 
Newtonian mechanics, that is the condition that the effective potential 
must be a minimum:
\begin{equation}
 \f{\partial^2 V_{\rm eff}(r,E,L)}{\partial r^2}|_{r=R} > 0.
\label{ep4}
\end{equation}
Eqs. (\ref{ep2}-\ref{ep4}) determine the ranges of $E,\, L, \,R$
allowed for SCOs in the KTN spacetime. At the 
ISCO, the one just on the verge of being unstable -- (eq. \ref{ep4})
becomes an equality. The last three equations are solved to obtain
the values of $E,\, L, \,R=r_{\rm ISCO}$ that characterize the orbit \cite{jia}.
Eqs. (\ref{ep2}) and (\ref{ep3}) were already solved in \cite{cc2} to 
obtain the energy ($E$) and angular momentum ($L$) of a test particle
moving in a circular orbit in the KTN spacetime. For direct orbits, 
we can write $E$ and $L$ as \cite{cc2, ryan} 
\begin{eqnarray}
 E=\f{r^{\f{1}{2}}(r^2-2M r-n^2)+a m^{\f{1}{2}}}
 {\left[(r^2+n^2)\left(r^3-3M r^2-3n^2r+M n^2+2a(m r)^{\f{1}{2}}\right)\right]^{\f{1}{2}}},
\label{E}
\end{eqnarray}

\begin{eqnarray}
L=\f{m^{\f{1}{2}} (r^2+a^2+n^2)-2a r^{\f{1}{2}}(M r+n^2)}
 {\left[(r^2+n^2)\left(r^3-3Mr^2-3n^2r+M n^2+2a(m r)^{\f{1}{2}}\right)\right]^{\f{1}{2}}}
\label{L}
\end{eqnarray}
respectively, where $m=M~(r^2-n^2)+2~n^2r$. We note that eqs. (\ref{E}--\ref{L})
reduce to eqs. (12.7.17--12.7.18) of \cite{st} for $n=0$, in case of the Kerr spacetime.
However, as the ISCO equation was also obtained in \cite{cc2} from the effective
potential formulation, we do not repeat it in this section.

\begin{figure}[h!]
 \begin{center}
\subfigure[KTN BH with $n_*=1$ and $a_*=1.20$ for $E=0.9$. ISCO is located at 
$r_{\rm ISCO}=3.6M$.]{\includegraphics[width=2.8in,angle=0]{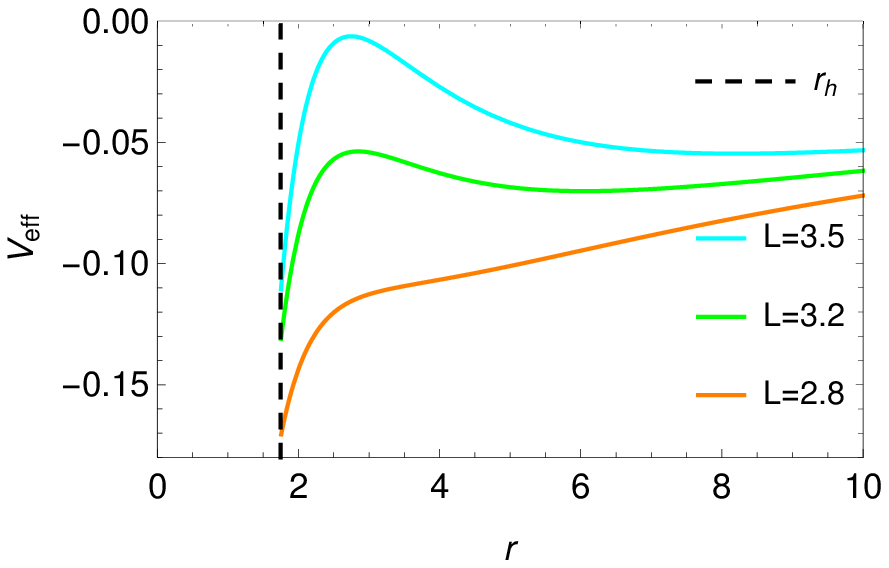}}
\hspace{0.05\textwidth}
\subfigure[KTN NS with $n_*=1$ and $a_*=1.45$ for $E=0.9$.
ISCO does not exist.]{\includegraphics[width=2.8in,angle=0]{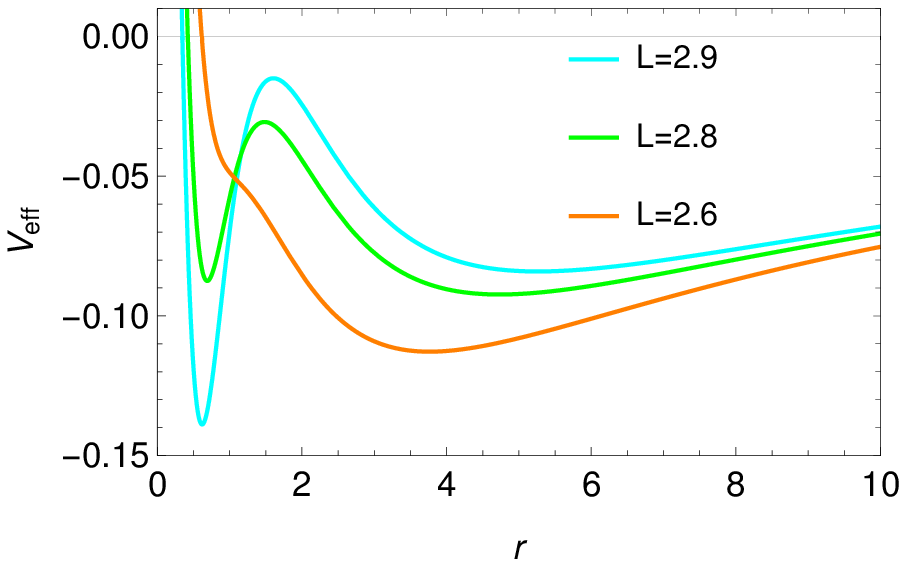}}
\hspace{0.05\textwidth}
\subfigure[Kerr BH with $a_*=0.90$ for $E=0.9$.
ISCO is located at $r_{\rm ISCO}=2.32M$.]{\includegraphics[width=2.8in,angle=0]{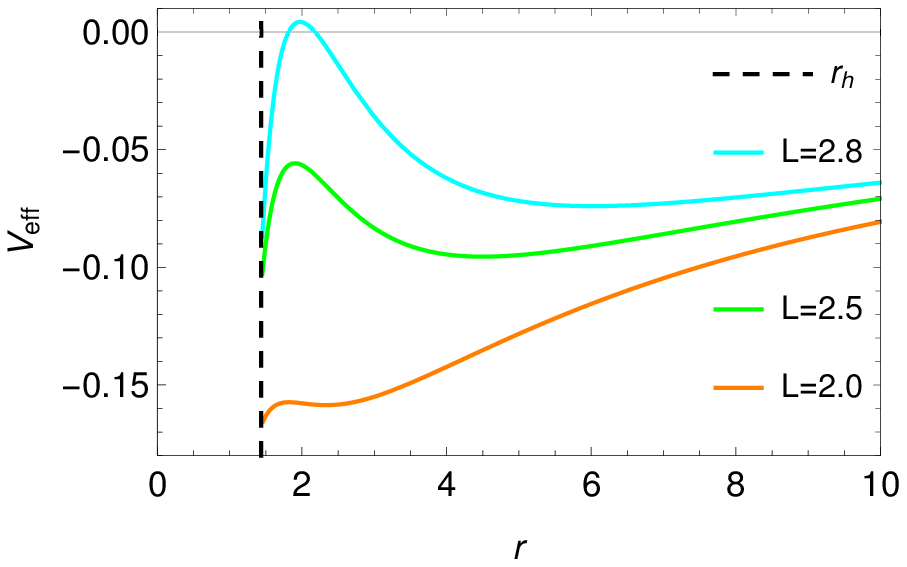}}
\hspace{0.05\textwidth}
\subfigure[Kerr NS with $a_*=1.05$ for $E=0.6$.
ISCO is located at $r_{\rm ISCO}=0.68M$.]{\includegraphics[width=2.8in,angle=0]{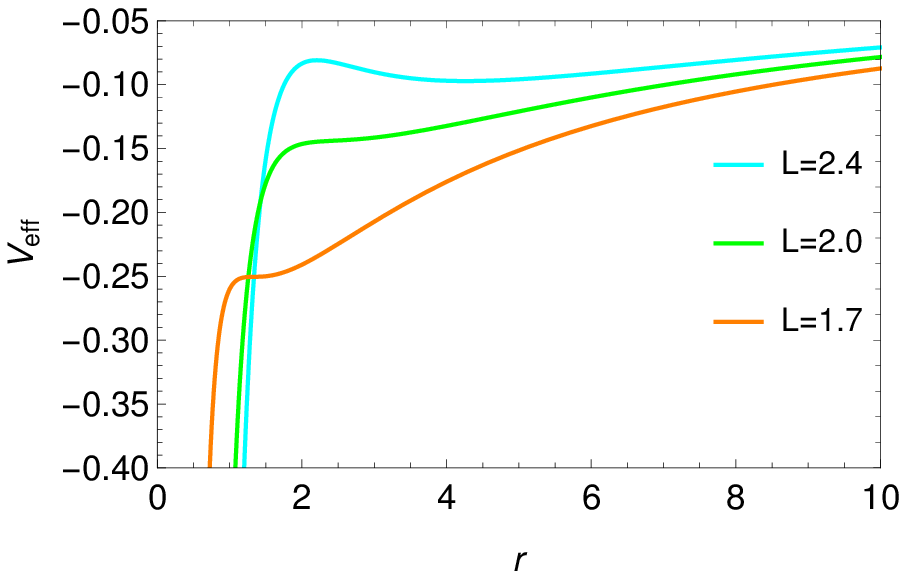}}
\caption{\label{V}$V_{\rm eff}$ vs $r$ (in `$M$') for various values of 
$L$ (in `$M$') with the fixed $E$. $r_h$ indicates the radius of horizon.
See section \ref{effp} for details.}
\end{center}
\end{figure}

The features of the effective potential curve of the KTN BH and KTN NS could be 
seen from panels (a) and (b) of figure \ref{V}, respectively. To compare it with Kerr
spacetime, we can follow panels (c) and (d) of the same figure. One can see from panel (b)
that the value of $V_{\rm eff}$ is finite at $r=0$ for a KTN NS and the value of $V_0$ can 
be calculated using eq. (\ref{v0}) but it diverges in the case of a Kerr NS (see panel (d)).
Here we should note that the ISCO radii of the Kerr BH and Kerr NS have been mentioned in 
the plots of panels (c) and (d) of figure \ref{V}, as these values are well-known. However,
the detailed studies of the locations of ISCOs in the KTN BH (panel a) and KTN NS (panel b)
will be discussed in section \ref{sco}. For the time-being, it could be noted (also clearly 
seen from the $V_{\rm eff}$ plots of panel (a)) that the ISCO occurs at $r_{\rm ISCO}=3.6M$ 
for the KTN BH with $n_*=1$ 
and $a_*=1.20$. However, one cannot determine the ISCO radius of the KTN NS with $n_*=1$
and $a_*=1.45$ from panel (b) of figure \ref{V}, as two local minima and one local 
maxima occur in this case. We discuss it in section \ref{sec:iscons}.

\subsection{ISCO equation from the expression of radial epicyclic frequency}\label{or}
One can deduce the ISCO equation directly from the expression of radial 
epicyclic frequency $(\O_r)$.
To derive the epicyclic frequencies one should investigate the small perturbations of a 
circular orbit. Perturbing the circular orbit with coordinate radius $r$, one can derive
the radial epicyclic frequency (see eq. (11) of \cite{don2} and the discussions below 
of that equation). However, from the above-mentioned analysis, it is known
that the square of the radial epicyclic frequency ($\O_r^2=0$) becomes zero at the 
ISCO \cite{don, don2, cpjcap}. No radial instability exists at the ISCO, whereas $\O_r^2$ becomes
negative for smaller radii. The negative sign of $\O_r^2$ also implies 
that there cannot be a radial oscillation for smaller radii than the ISCO.
It would be useful to mention here the general analytical expression of the 
epicyclic frequencies ($\o_x$) in the stationary and axisymmetric spacetime,
which is expressed as (see eq. (27) of \cite{abkl}),
\begin{eqnarray}
 \o_x^2 = \left(\f{\p^2\mathcal{U_{\rm eff}}}{\p X^2} \right).
 \label{ger}
\end{eqnarray}
Here $\mathcal{U_{\rm eff}}$
is the general expression of effective potential in a stationary and axisymmetric 
spacetime with $dX^2=g_{xx} dx^2 > 0$ being the proper length 
in the $x$ direction ($x$ denotes either radial $r$ or polar
angle $\th$ coordinate). Moreover, as $\o_x$ are measured with respect to the 
proper time of a comoving observer, after dividing it by the squared redshift factor, 
one can obtain the observed epicyclic frequencies ($\O_x$) at infinity
(see eqs. 32-33 of \cite{abkl} and related discussions there for details). 
Now, it is clear from eq. (\ref{ger}) that the sign of $\o_r^2$ (and also $\O_r^2$) is
correlated with the sign of the second derivative of effective
potential. However, $\O_r^2$ is positive for 
a SCO and we will show in this paper that all derived SCOs in the 
KTN spacetime cannot be accessible for a test particle. Rather, some SCOs 
cannot exist in reality and the innermost SCO is determined from the
remaining SCOs. 

Before going into the more detail, we should first clarify that
we mainly focus on the equatorial circular orbits in this paper. Therefore,
we can directly use the expressions of three fundamental
frequencies which have already been deduced for KTN spacetime in \cite{cbgm}
using the general formulation derived by Ryan \cite{ryan}. These are orbital
frequency ($\O_{\phi}$) or the Kepler frequency 
\begin{eqnarray}
\O_{\phi}=\pm \f{m^{\f{1}{2}}}{r^{\f{1}{2}}~(r^2+n^2) \pm a~m^{\f{1}{2}}},
\label{kktn}
\end{eqnarray}
where $m=M~(r^2-n^2)+2~n^2r$, and the radial ($\O_r$) and vertical ($\O_{\th}$)
epicyclic frequencies \cite{cbgm}
\begin{eqnarray}\nonumber
\O_r &=& \pm \f{1}{(r^2+n^2)\left[r^{\f{1}{2}}~(r^2+n^2) \pm a~m^{\f{1}{2}}\right]}.              
\left[M(r^6-n^6+15n^4r^2-15n^2r^4)-16n^4r^3 \right. \nonumber
\\
&& \left. -2M^2r (3r^4-2n^2r^2+3n^4) 
\pm 8ar^{\f{3}{2}}m^{\f{3}{2}}+a^2\left\{M(n^4+6n^2r^2-3r^4)-8n^2r^3 \right\}\right]^{\f{1}{2}},  \nonumber
\\
\label{rktn}
\end{eqnarray}

\begin{eqnarray}\nonumber
\O_{\th} &=& \pm \f{1}{(r^2+n^2)\left[r^{\f{1}{2}}~(r^2+n^2) \pm a~m^{\f{1}{2}}\right]}. 
\left[M(r^6-n^6+15n^4r^2-15n^2r^4)  +2n^2r (3r^4-2n^2r^2+3n^4) \right.
\\
&& \left. + 16M^2n^2r^3 \mp 4ar^{\f{1}{2}}m^{\f{1}{2}}(n^2+Mr)(n^2+r^2)
-a^2\left\{M(n^4+6n^2r^2-3r^4)-8n^2r^3 \right\}\right]^{\f{1}{2}}
\label{vktn}
\end{eqnarray}
respectively. We note that the upper sign is applicable for the direct orbits and
the lower sign is applicable for the retrograde orbits.

Now, setting eq. (\ref{rktn})
equal to zero, we obtain the so-called ISCO equation :
\begin{eqnarray}\nonumber
&& M(r^6-n^6+15n^4r^2-15n^2r^4)-2M^2r (3r^4-2n^2r^2+3n^4)-16n^4r^3 
\\
 && \pm 8ar^{\f{3}{2}}m^{\f{3}{2}} +a^2\left\{M(n^4+6n^2r^2-3r^4)
-8n^2r^3 \right\}=0.
\label{isco}
\end{eqnarray}
One can check that eq. (\ref{isco}) reduces to the ISCO equation in Kerr spacetime
\cite{ch} for $n=0$  :
\begin{eqnarray}
r^2 -6Mr \pm 8ar^{\f{1}{2}}M^{\f{1}{2}}-3a^2=0
\label{iscokerr}
\end{eqnarray}
and obtain only one positive real root of this equation for each corresponding value of $a_*$ 
(see figure 7 of \cite{ckp}), which is considered as the radius of the 
ISCO for that particular value of $a_*$. This is true for all values of $a_*$ whether it indicates
a BH ($-1 \leq a_* \leq 1$) or a NS ($ a_* > 1,\,\, a_* < -1$).
One intriguing feature is that eq. (\ref{isco}) gives more than one positive real root
for one class of combinations between $a_*$ and $n_*$. More interestingly, eq. (\ref{isco}) does not
give any positive real root for another class of combinations between $a_*$ and $n_*$, 
which indicates that the so-called ISCO does not exist for these cases. We will discuss all
these cases as we proceed.

\section{Stable circular orbits in KTN spacetime}\label{sco}
In this section, we study the properties of the circular geodesics of a test particle,
occurred in the KTN spacetime and find the locations of the ISCO as well as the other 
SCOs. We divide it into two subsections. 

\subsection{Location of ISCO in KTN black hole}\label{iscobh}
The ISCO equation (eq. \ref{isco}) cannot be solved analytically, but one can 
solve it numerically and obtain the solutions for all possible combinations of
$a_*$ and $n_*$. For the non-extremal KTN BH, we obtain only one positive real
root of eq. (\ref{isco}) outside the horizon $r_h = M(1+\sqrt{1+n_*^2-a_*^2})$.
\footnote{In this manuscript, we do not consider the roots of any equation inside 
the horizon, as this is irrelevant for an accreting
black hole from the observational point of view. We focus only those solutions which
occur outside as well as on the horizon.}
Hence, we can conclude that the ISCO always occurs
outside of the horizon : $r_{\rm ISCO} > r_h$ for all possible combinations of
$a_*$ and $n_*$ which represent the non-extremal KTN BHs. This is expected.

In an extremal KTN BH, ISCO equation for the direct orbits (eq. \ref{isco}) reduces to 
\begin{eqnarray}\nonumber
&& M(r^6-n^6+15n^4r^2-15n^2r^4)-2M^2r (3r^4-2n^2r^2+3n^4)-16n^4r^3 
\\
 && + 8(M^2+n^2)^{\f{1}{2}}r^{\f{3}{2}}m^{\f{3}{2}} +(M^2+n^2)\left\{M(n^4+6n^2r^2-3r^4)
-8n^2r^3 \right\}=0
\label{iscox}
\end{eqnarray}
and it is satisfied by $r_1=M$ which is independent of $n_*$. This also means that this solution
is satisfied by all values of $n_*$ and coincides with the horizon $r_1=r_h=M$. We note that 
the ISCO equation (eq. \ref{iscokerr}) for an extremal Kerr BH  is satisfied by $r=M$ 
and therefore, ISCO occurs at $r=M$. One intriguing feature is that eq. (\ref{iscox})
is also satisfied by another positive real root ($r_2$) which occurs outside (i.e., $r_2 > M$)
or inside (i.e., $r_2 < M$) the horizon depending on the value of $n_*$.
For $0 < n_* < 0.577$, $r_2$ occurs inside the horizon whereas it occurs outside the horizon 
for $n_* > 0.577$, and $r_2$ coincides with the horizon ($r_1=r_2=M$) for $n_* \approx 0.577$.
As the orbits inside the horizon are unfeasible for an accreting BH, one should consider $r_1=M$ as the `physical 
ISCO' for an extremal KTN BH with $n_*$ value of the following range, $n_* : 0 < n_* \lesssim 0.577$.
However, referring to section \ref{or},
we plot the radial epicyclic frequency ($\O_r$) for a few values of $n_*$.
\begin{figure}
   \begin{center}
\includegraphics[width=4in]{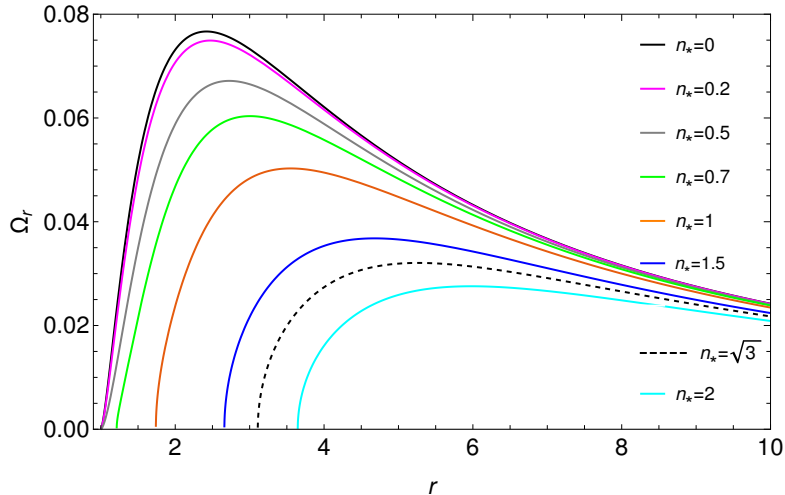}
\caption{\label{frefx}
$\O_r$ (in `$M^{-1}$') vs $r$ (in `$M$') for different
values of $n_*$ of the extremal KTN BHs. X-axis starts from the event horizon $r_h=M$.
Solid black curve ($n_*=0$) stands for the extremal Kerr BH and the ISCO is located at 
$r_{\rm ISCO}=M$ in this case. Each curve (except the solid black, magenta and gray curves) 
touches the X-axis at $r=r_2$ which is considered as the `physical ISCO' for an extremal 
KTN BH. Magenta and gray curves touch the X-axis at $r_1=M$ which is considered
as the ISCO for these two cases. See section \ref{iscobh} for details.}
     \end{center}
\end{figure}
Except solid black, magenta and gray curves, all the curves of figure \ref{frefx}
indicate the positions of $r_2$ where $\O_r$ vanishes. 
We note that solid black curve shows the position of ISCO for an extremal Kerr BH
whereas magenta and gray curves show the position of $r_1=M$, as is discussed above.
We confirm that $\O_r$ also vanishes at $r_1=M$ for 
all these cases. It suggests that no radial instabilities should be found at $r_1$ and
$r_2$ (see the related discussions in section \ref{or}), in principle. 
However, as the values of $\O_r^2$ become negative between $r_1$ and 
$r_2$ for the extremal KTN BHs with $n_* > 0.577$, SCOs do not exist. Now, question
is which should be the ISCO ($r_1$ or $r_2$) 
in a realistic astrophysical situation for the extremal KTN BHs with $n_* > 0.577$? 
The answer of this question, i.e., the location of `physical ISCO' 
for the extremal KTN BH could not be found from the effective potential plot, shown 
in panel (a) of figure \ref{Vx} (effective potential in the extremal Kerr BH is shown
in panel (b) only for the comparison).

\begin{figure}
\centering
\subfigure[KTN extremal BH with $n_*=1$ \& $a_*=\sqrt{2}$ for $E=0.8$.
`Physical ISCO' is located at $r_2=1.75M$]{\includegraphics[width=2.8in,angle=0]{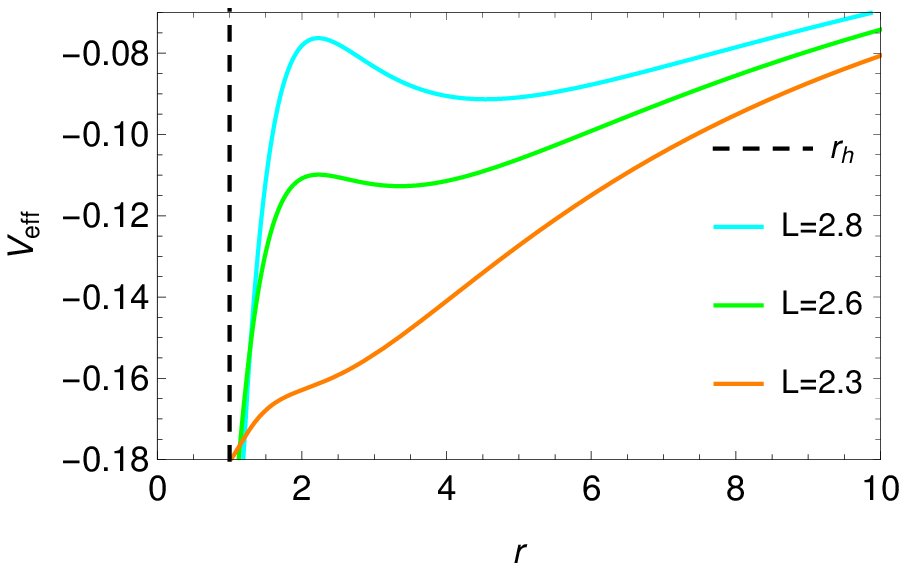}}
\hspace{0.05\textwidth}
\subfigure[Kerr extremal BH with $a_*=1$ for $E=0.7$. 
ISCO is located at $r_{\rm ISCO}=M$]{\includegraphics[width=2.8in,angle=0]{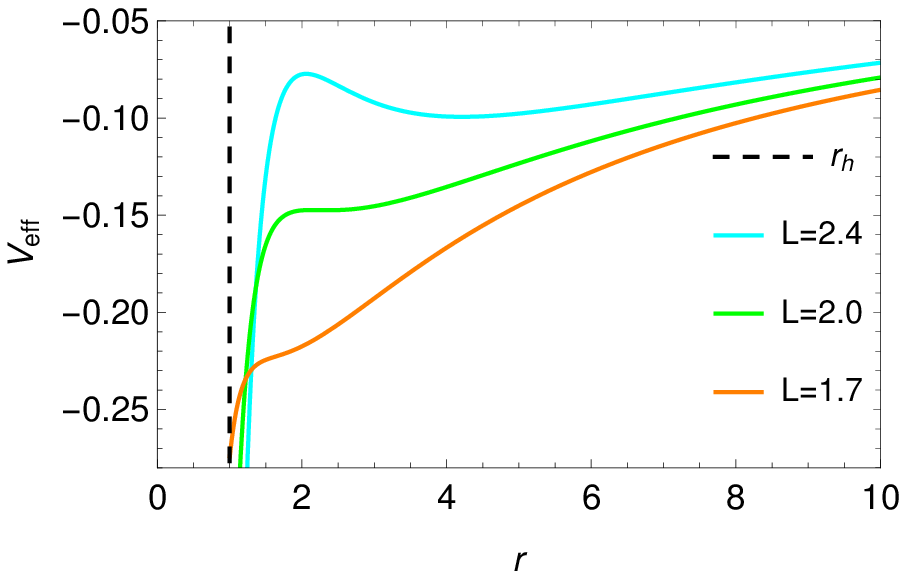}}
\caption{\label{Vx}$V_{\rm eff}$ vs $r$ (in `$M$') for various values of 
$L$ (in `$M$') with the fixed $E$. $r_h$ indicates the radius of horizon.
See section \ref{iscobh} for details.}
\end{figure}

Therefore, to find the answer of this important question, we calculate the energy
at $r_1$ and $r_2$ using eq. (\ref{E}). We find that $E$ comes out 
as less than $1$ in all the orbits which are in the range of $r \geq r_2$. This 
indicates that all these orbits are accessible by a test particle and these 
are realistic for an astrophysical purpose. Now, if we calculate the energy at $r_1$,
we obtain
{\footnote{For comparison, we note that the energy and angular momentum
of an extremal Kerr BH can be expressed as \cite{jh}
\begin{eqnarray}\nonumber
 E_{\rm ISCO}^{\rm Kerr}=\f{1}{\sqrt{3}} \,\,\,\,\,\, {\rm and} \,\,\,\,\,\,
 L_{\rm ISCO}^{\rm Kerr}=\f{2M}{\sqrt{3}}
 \label{lek}
\end{eqnarray}
at $r=M$ which is basically the ISCO for the extremal Kerr BH or $a_*=1$.}}
\begin{eqnarray}
 E_{r=r_1=M}={\sqrt\f{1+n_*^2}{3-n_*^2}} 
 \label{ex}
 \end{eqnarray}
and the corresponding angular momentum is
 \begin{eqnarray}
 L_{r=r_1=M}=\f{2M(1+n_*^2)}{\sqrt{3-n_*^2}}.
 \label{lx}
\end{eqnarray}
At this point, we should note that all circular orbits are not bound. 
The orbit of a test particle with energy $E=1$ is regarded \cite{ch} as the 
marginally bound orbit ($r_{\rm mb}$). Therefore, the bound circular orbits 
exist for $r > r_{\rm mb}$ with $E < 1$, whereas unbound circular orbits exist for $r < r_{\rm mb}$ 
with $E > 1$ \cite{st}. If an infinitesimal outward perturbation is given, a test particle in an 
unbound orbit escapes to infinity on an asymptotically hyperbolic trajectory. 
However, in realistic astrophysical problems, particle infall from infinity is very nearly 
parabolic, and any parabolic trajectory, penetrating to $r < r_{\rm mb}$, must plunge 
directly into the collapsed object \cite{st}.  
Now, we can see from eq. (\ref{ex}) that $E \geq 1$ at $r=r_1$ for $1 \leq n_* < \sqrt{3}$,
{\footnote{$E \rightarrow \infty$ for $n_* \rightarrow \sqrt{3}$, which represents the CPO.
See section \ref{cpobh} for details. $r_1$ is completely meaningless for $n_* > \sqrt{3}$ as 
$E$ becomes imaginary.}}
and $E < 1$ for $n_* < 1$. Therefore, the single orbit which occurs at $r=r_1=M$
could not be important in a realistic astrophysical situation for $1 < n_* < \sqrt{3}$, as the 
particle will plunge directly into the black hole from the $r_2$ orbit.
Therefore, we should identify $r=r_2$ as the ISCO in this case. 

For $0.577 < n_* \leq 1$, the $r=r_1$ orbit must be a SCO in principle 
and it should exist but it could not be relevant for the astrophysical purpose. 
Let us first consider
that an extremal KTN BH with $0.577 < n_* \leq 1$ is accreting matter from a distant source. The matter 
generally follows the SCOs and the accretion disk exists till $r=r_{\rm ISCO}$.
The accreting matter free-falls towards the BH at $r < r_{\rm ISCO}$, as no SCOs exist there.
It is clear from figure \ref{frefx} that SCOs exist for $r \geq r_2$ (as 
$\O_r^2 \geq 0$) and no SCOs exist in this range $r_1 < r < r_2$ (as 
$\O_r^2 < 0$). In such a situation, one
cannot expect that the accreting matter will form a `single-ring' disk on the event horizon,
i.e., $r=r_1=M$ after a free-fall from $r > r_2$ to $r = r_1$ which basically represents
the boundary of the BH. Therefore, one should 
consider $r=r_2$ as the `physical ISCO' for all practical purposes in the case of
an extremal KTN BH with $n_* > 0.577$.
Figure \ref{frefx} also reveals that the radius of ISCO ($r_2$ in this case)
increases with increasing the value of $n_*$ of the extremal KTN BHs with 
$n_* > 0.577$. However, as we have already discussed, $r_1$ should be the `physical ISCO' for 
an extremal KTN BH with $0 < n_* \lesssim 0.577$.

\subsection{Location of ISCO in KTN naked singularity}\label{sec:iscons}
\begin{figure}
\includegraphics[width=4.8in]{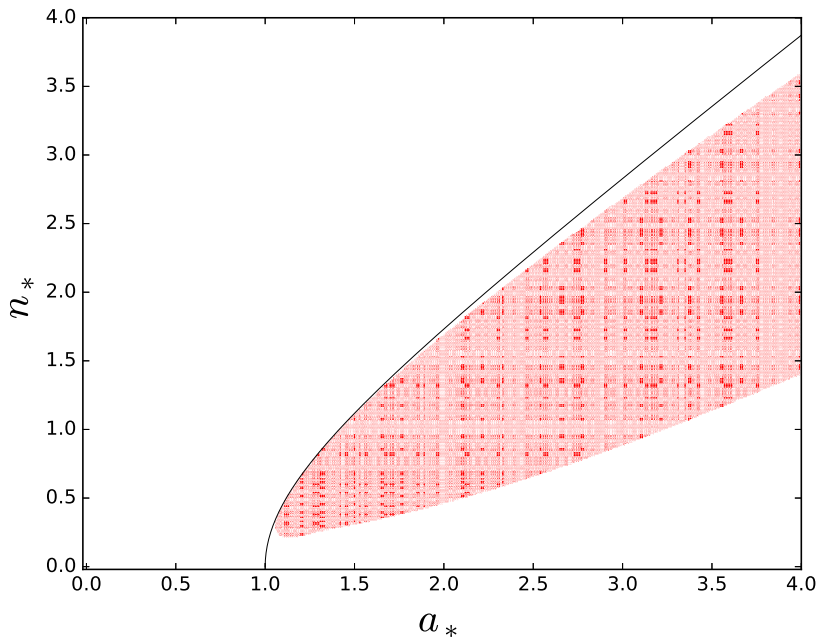}
\centering
 \caption{\label{noisco}NUT parameter ($n_*$) versus Kerr parameter ($a_*$) space, which
 is divided into a KTN BH and KTN NS region by the thin black curve. The $a_*$ and $n_*$ values of
 each point in the thin black curve also indicates an extremal KTN BH.
 Solving eq. (\ref{isco}) one cannot obtain a positive real root, 
 for those combinations of $a_*$ and $n_*$ which are fallen in the `red-coloured' region, whereas
at least one positive real root can be obtained, if one choses the combinations of 
$a_*$ and $n_*$ from the `white-coloured' region. One should note that the `red-coloured' region
does not include any KTN BH. Because, solving eq. (\ref{isco}) one can always obtain 
at least one positive real root outside or on the horizon for all KTN BHs.
See sections \ref{iscobh} and \ref{sec:iscons} for details.}
\end{figure}

As the so-called ISCO equation (eq. \ref{isco}) could not be solved analytically, 
we solve it numerically and obtain three classes of solution. For one 
class of solution, we obtain two positive real roots ($r_1$ and $r_2$) of eq. (\ref{isco}). For the second
class, one can obtain one (double root) positive real root, and for another class, 
no positive real roots are found. Basically, in the second class of solution, two positive real roots become
equal, i.e., $r_1=r_2$ for some special combinations of $a_*$ and $n_*$. Our result has been depicted in figure \ref{noisco}. It shows
that one cannot obtain any positive real root of eq. (\ref{isco}) for those combinations of $a_*$ and 
$n_*$, which are fallen in the red region, whereas at least one positive real root 
can be obtained, if one choses the values of $a_*$ and $n_*$ from the white region.
In this sense, the ISCO should not exist for the KTN NSs which are in the `red-coloured' region 
and SCOs can exist everywhere in this spacetime. Here, we should note that,
considering a toy model of a static spherically symmetric
perfect fluid interior with a singularity at the origin, it was found \cite{jmn}
that the SCOs exist everywhere (till $r\rightarrow 0$).
However, our case is completely different here, which we will be discussing
in this section considering the various scenarios. Let us first consider an example.

\begin{figure}
\centering
\subfigure[Different of $\O_r$ curves shows that the locations of SCOs are highly affected
with a slight 
change in the Kerr/NUT parameter values. As an example, $a_*$ changes its value with a 
fixed $n_*$ in this particular figure.]{\includegraphics[width=2.8in,angle=0]{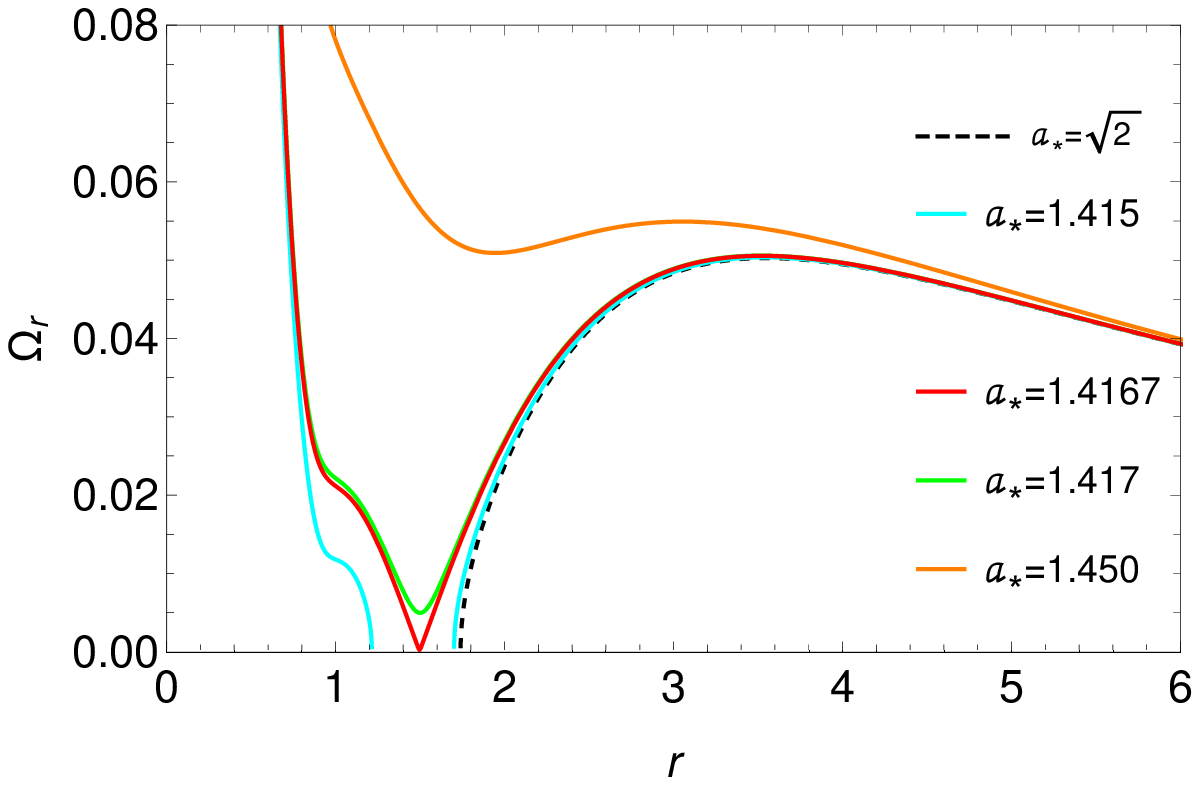}}
\hspace{0.05\textwidth}
\subfigure[Zoomed version of panel (a) with range $0.4 \leq r/M \leq 0.6$
is shown for clarity. $\O_r$ curves are discontinued in the 
region : $r < 0.414M$.]{\includegraphics[width=2.8in,angle=0]{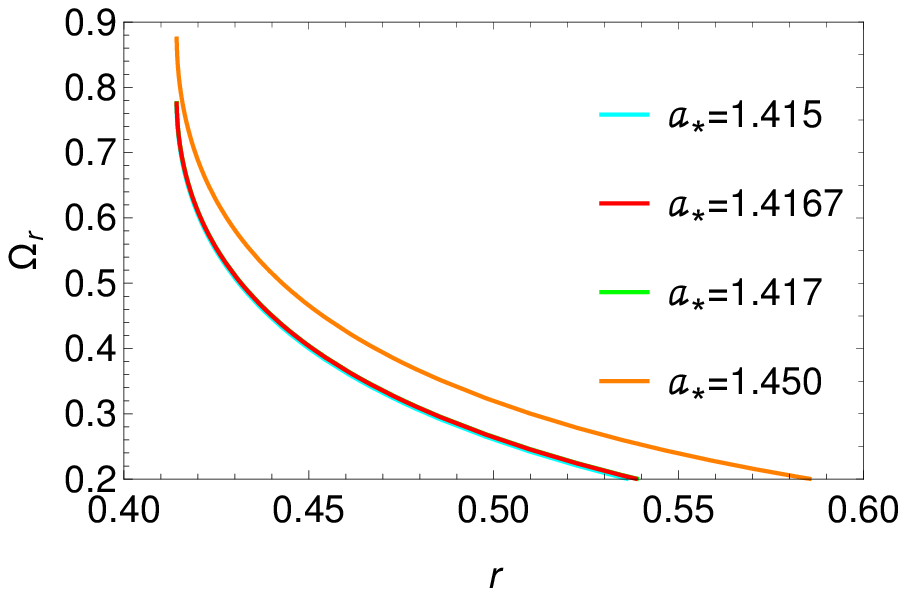}}
\caption{\label{ref}$\O_r$ (in `$M^{-1}$') vs $r$ (in `$M$') for various values of $a_*$
 with $n_*=1$. Panel (a) also clearly shows that three classes of solution 
 (two, one and zero positive roots)
of the ISCO equation (eq. \ref{isco}) exist for KTN NS. The orange and green curves show that 
 one cannot obtain a positive real root by solving of eq. (\ref{isco}) as $\O_r$ does not 
 vanish in any orbit. But, one can obtain two positive real roots for the cyan curve
 and one positive real root for the red curve. Actually, the last one is the double root,
 i.e., two positive real roots coincide at $r/M=1.5$ for $a_*=1.4167$, which is clear 
 from this figure. See section \ref{sec:iscons} for details. For completeness,
 the black dashed curve is added, which stands for an extremal KTN BH. We have
 already discussed on it in section \ref{iscobh}}.
\end{figure}

In figure \ref{ref}, we have plotted $\O_r$ for different values of $r$. It shows that 
the black dashed curve which stands for the extremal KTN BH ($n_*=1, a_*=\sqrt{2}$) 
touches the X-axis at $r/M=1.75$, which means that eq. (\ref{isco}) has a positive real
root at $r/M=1.75$. It has already been recognized as $r_2$ or the `physical
ISCO' for this case (see also the solid orange curve in figure \ref{frefx} and the related 
discussion in section \ref{iscobh}). Now, if we slightly increase the value of $a_*$ from
$\sqrt{2}$ to $1.415$ fixing the value of $n_*=1$, the event horizon vanishes and we should
consider it as a KTN NS. In this case, the solid cyan curve of figure \ref{ref} touches
the X-axis twice at $r_1/M=1.2$ and $r_2/M=1.7$ respectively. This means that eq. (\ref{isco}) 
has two positive real roots at $r_1$ and $r_2$. 
If we again increase the value of $a_*$ from $a_*=1.415$ to $a_*=1.4167$, the two roots 
become equal and coincide at $r_1=r_2=1.5$. This means, that one can obtain only 
one positive real root in this particular case. This is shown by the solid red curve 
in figure \ref{ref}.
The green and orange curves do not touch the X-axis, which means that one cannot obtain 
any positive real root of eq. (\ref{isco}) for these two curves,
i.e., for $n_*=1$ \& $a_*=1.417$ and $n_*=1$ \& $a_*=1.45$, respectively. As we have discussed
in section \ref{or},
one can think that SCOs exist for all those values of $r$ which gives the positive $\O_r$
values in principle. It can easily 
be seen that the SCOs exist in the outer branch ($r_2 \leq r < \infty$) of the cyan curve
but for the inner branch ($r \leq r_1$) it might not be true always. We will discuss it in 
the next two sections.
A close look of panel (b) (the zoomed version of panel (a)
of figure \ref{ref}) reveals that the cyan, red, green and orange curves (which are
for KTN NSs) do not continue to $r \rightarrow 0$. Rather, all these 
curves are discontinued after arriving at a particular 
orbit of radius $r=R_f$. Mathematically, this discontinuation indicates that the value of $\O_r$ 
becomes imaginary in the region : $r < R_f$, which is unphysical. One can check that 
the Kepler frequency (eq. \ref{kktn}) vanishes 
at the orbit $r=R_f$, which is unexpected. As far as we
know, this special feature has not been seen in other spacetimes until now. Therefore, we should
discuss it in detail in the next section. Side by side, we also continue our discussion 
on SCOs in the later sections.

\subsubsection{Digression 1 : Forbidden region for a test particle 
moving in a circular geodesic}\label{fr}
\begin{figure}
   \begin{center}
\includegraphics[width=3.3in]{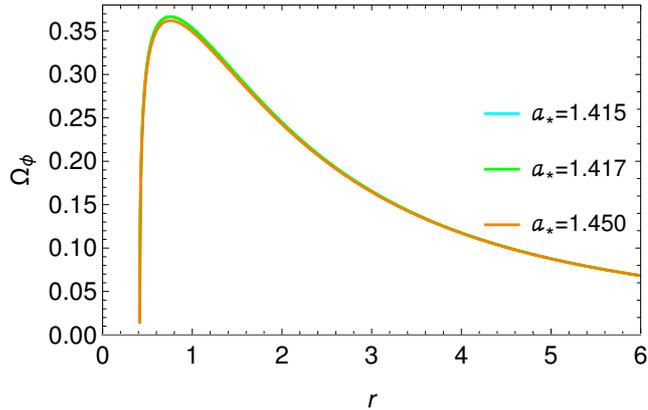}
     \caption{\label{kep}$\O_{\phi}$ (in `$M^{-1}$') vs $r$ (in `$M$') for various
     values of $a_*$ with $n_*=1$. One intriguing feature is that $\O_{\phi}$ vanishes 
     at $r=0.414M$. See section \ref{fr} for details.}
     \end{center}
\end{figure}
\begin{figure}
   \begin{center}
\includegraphics[width=3.3in]{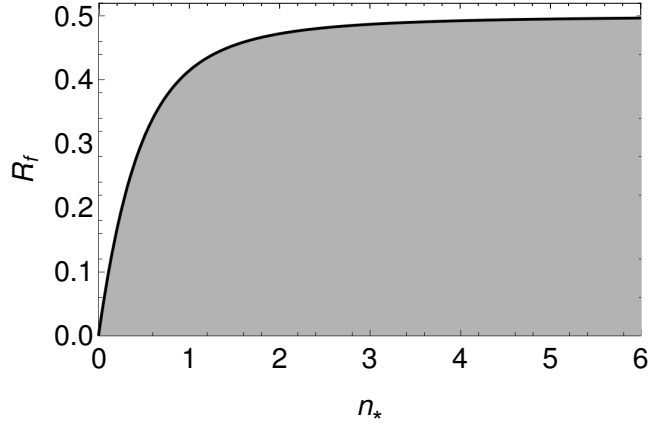}
     \caption{\label{frf}$R_f$ (in `$M$') vs $n_*$, where Gray region indicates
     the ``Forbidden Region'' : $r < R_f$. The solid black curve indicates the boundary 
  values of the forbidden region, for each corresponding value of $n_*$.
  $\O_{\phi}$ vanishes on the solid black curve. See section \ref{fr} for details.}
     \end{center}
~\label{fref}
\end{figure}

Figure \ref{kep} shows that the Kepler frequency of a test particle which moves in a circular 
geodesic, increases at first, attains a peak value at $r=r_p$ and then vanishes at a particular 
orbit of radius $r=R_f$. The value of $R_f$ can be
obtained setting the numerator of eq. (\ref{kktn}) as zero and it gives
\begin{eqnarray}
 R_ f= M \left(n_* \sqrt{1+n_*^2}-n_*^2 \right).
 \label{rf}
\end{eqnarray}
Therefore, a test particle moving in a circular geodesic cannot continue its motion 
at $r < R_f$
as its angular velocity is zero. We call it as the forbidden region. As the 
angular velocity vanishes at $r=R_f$, the test particle should free-fall towards the
central object. In the next sections, we will show that the test particle cannot even
continue its stable circular motion till $r\rightarrow R_f$, in a realistic situation.
However, we can see from 
eq. (\ref{rf}) that the forbidden region is fully controlled by the gravitomagnetic monopole $n_*$
and Kerr parameter ($a_*$) has no influence on it. Therefore, this region is absent in the Kerr 
spacetime, i.e, $R_f=0$ (see figure \ref{frf}). 
If $n_*$ increases from $0$ to a higher value, the radius of 
the forbidden region increases primarily but it cannot be greater than $M/2$. 
A short calculation reveals that 
\begin{eqnarray}
 \f{R_f}{M}|_{n_* >> 1} \approx n_*^2 \left(1+\f{1}{2n_*^2}\right)-n_*^2 = \f{1}{2}
\end{eqnarray}
for higher values of $n_*$.
It means that for $ n_* >> 1$, the higher order terms could  be neglected and the radius of
the forbidden region becomes almost a constant : $M/2$ which is also clear from 
figure \ref{frf}. The gray region indicates the forbidden region for various values of $n_*$. 
As $R_f$ is always less than $ r_h$, i.e., $R_f < r_h$, it remains hidden inside the horizon
and this region could only be meaningful in case of a KTN NS. For a particular value of $n_*$, 
the Kepler frequency not only vanishes at $R_f$ in the strong gravity regime 
{\footnote{Though this is completely a new thing but not unusual in the strong 
gravity regime. Note that the orbital plane precession or the Lense-Thirring precession
can also vanish (see figure 2 of \cite{cbgm}) in the strong gravity regime in case of a KTN 
BH \cite{cm,cc,cp}, which does not occur for a Kerr BH \cite{ckp}.}},
but the other two fundamental frequencies (radial
and vertical epicyclic frequencies) are also discontinued there, acquiring the finite
values. This implies that the circular motion become frozen in the region $r < R_f$
and cannot be accessed by a test particle with its circular geodesic motion.
Mathematically, all three fundamental frequencies $(\O_{\phi},\O_{\th}$ and $\O_{r})$
become imaginary at $r < R_f$.

However, differentiating eq. (\ref{kktn}) with respect to $r$ and
setting it to zero :
\begin{eqnarray}
 \f{d \O_{\phi}^{\rm KTN}}{dr}=\frac{-8 n^2 r^3+M \left(n^4+6 n^2 r^2-3 r^4\right)}
 {2 \sqrt{rm} \left[\sqrt{r}(r^2+n^2)+a\sqrt{m}\right]^2}|_{r=r_p}=0 
\end{eqnarray}
one can obtain the radius of the peak ($r=r_p$) 
where the Kepler frequency acquires the maximum value
\begin{eqnarray}\nonumber
r_p&=& \frac{M n_*}{3} 
\left[\left(6+8 n_*^2+3\left(1+n_*^2\right)^{1/3}+\frac{2 n_* \left(9+8 n_*^2\right)}
{\left[3+4 n_*^2-3(1+n_*^2)^{1/3} \right]^{\f{1}{2}}} \right)^{\f{1}{2}} \right.
\\
&& \left. -\left(3+4 n_*^2-3(1+n_*^2)^{1/3}\right)^{\f{1}{2}} -2n_*\right].
\label{rp}
\end{eqnarray}
Eq. (\ref{rp}) shows that the value of $r_p$ also does not depend on $a_*$ and 
therefore the similar situation cannot arise in case of a Kerr NS by any means.
\\

{\bf Conclusion drawn from this discussion :} The above discussion compels us to 
conclude that the SCOs do not exist in the region: $r \leq R_f$ for all those four
NS curves of figure \ref{ref}.

\subsubsection{Digression 2 : Angular velocity of the test particle inside the ergoregion}\label{av}
Now, the question is : can the SCOs exist in the range $R_f < r \leq r_1$ for the cyan curve
and $R_f < r < \infty$ for the red, green and orange curve of figure \ref{ref}(a)?

Here, we recall eq. (\ref{ergo}) where $r_e$ represents the boundary of the ergoregion.
Though the horizon does not exist for the KTN NS, the ergoregion remains there, and its radius becomes
\begin{eqnarray}
r_{e}|_{\th=\pi/2}=M \left(1 + \sqrt{1+n_*^2} \right).
 \label{ergo90}
\end{eqnarray}
For KTN spacetime it depends on the value of $n_*$ which is seen from eq. (\ref{ergo90}).
We note that the radius of ergoregion at the equatorial plane
remains same ($r_e|_{\th=\pi/2}=2M$) for all values of $a_*$,
whether it is a Kerr BH or a Kerr NS \cite{ckj}. However, it is well-known that it is impossible 
to stay fixed inside the ergoregion with some arbitrary velocities. 
In general, it is possible for an observer (or the test particle) to fix at a point $(r, \th)$ 
inside the ergoregion, if it rotates (prograde only) inside the ergoregion
\cite{jh,mtw} with the angular velocity $\O$ : $\O_-< \O < \O_+$. This range is 
determined by using eqs. (21-23) of \cite{ckp} :
\begin{eqnarray}
 \O_{\pm} = \frac{2 a \left(n^2+M r\right) \pm \left(n^2+r^2\right) 
 \sqrt{\d}}{\left(n^2+r^2\right)^2+a^2 \left[3 n^2+r(2 M+r)\right]}.
 \label{opm}
\end{eqnarray}
Therefore, inside the ergoregion, a test particle can rotate in an SCO with the 
Kepler frequency $\O_{\phi}$, if and only if it satisfies the following condition :
\begin{eqnarray}
 \O_- < \O_{\phi} < \O_+.
 \label{C}
\end{eqnarray}

\begin{figure*}
\centering
\subfigure[Kerr BH with $a_*$=0.99]{\includegraphics[width=2.8in,angle=0]{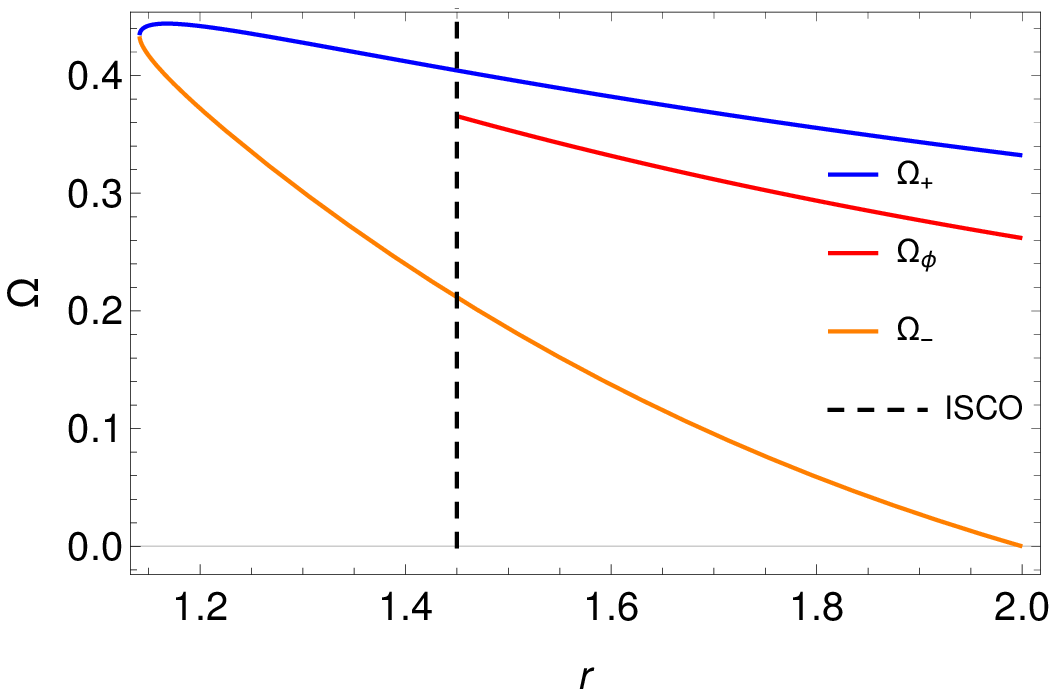}}
\hspace{0.05\textwidth}
\subfigure[Kerr NS with $a_*$=1.01]{\includegraphics[width=2.8in,angle=0]{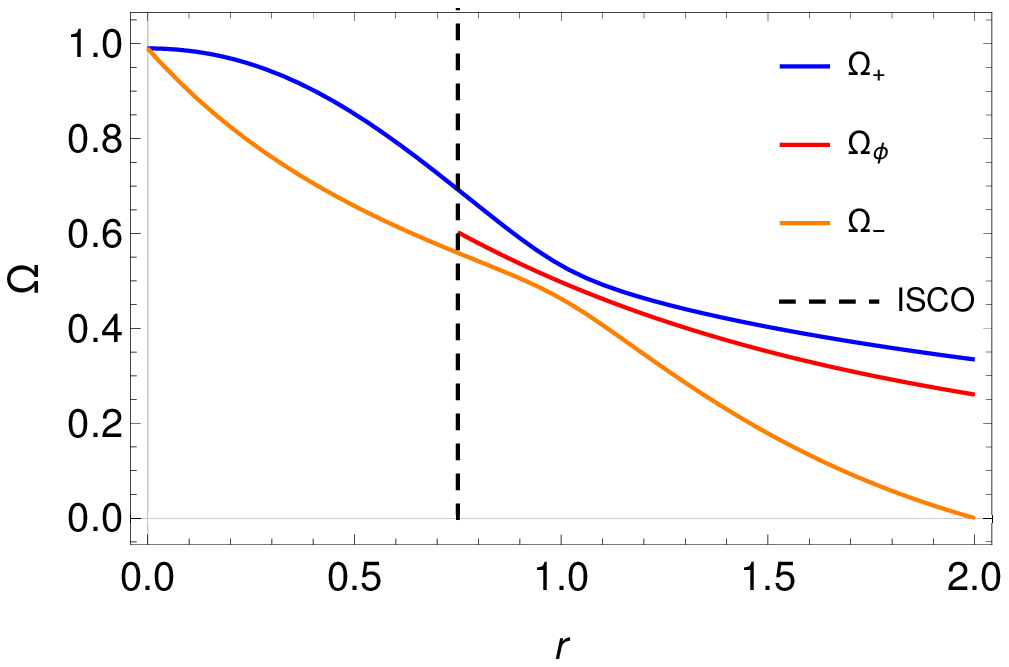}}
\hspace{0.05\textwidth}
\subfigure[KTN BH with $n_*=1$ and $a_*=1.4$]{\includegraphics[width=2.8in,angle=0]{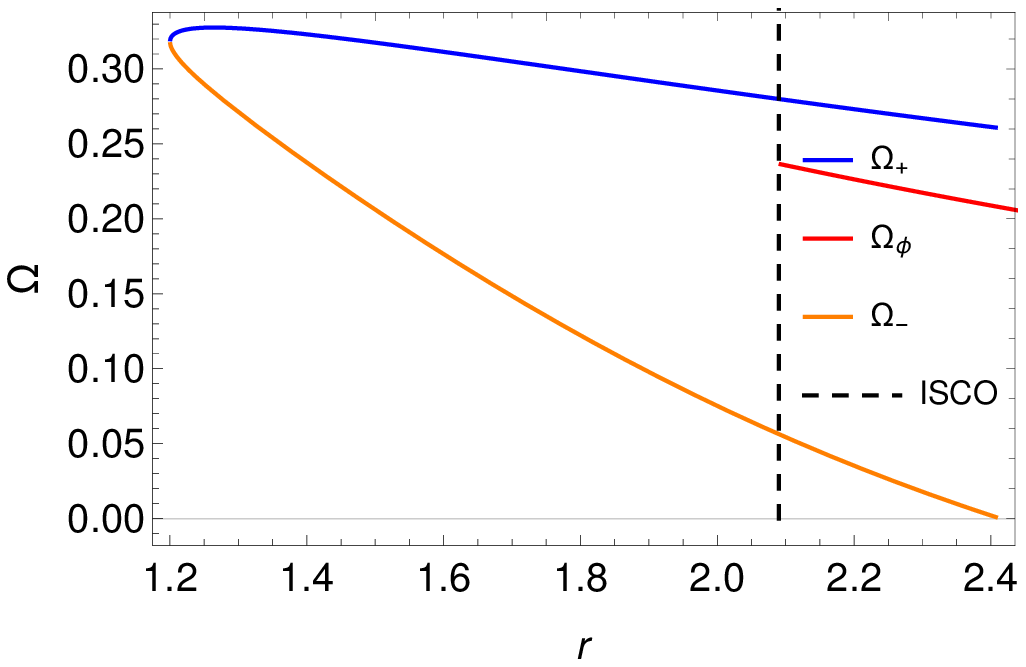}}
\hspace{0.05\textwidth}
\subfigure[KTN NS with $n_*=1$ and $a_*=1.415$ (corresponds to the cyan curve of figure \ref{ref})]{\includegraphics[width=2.8in,angle=0]{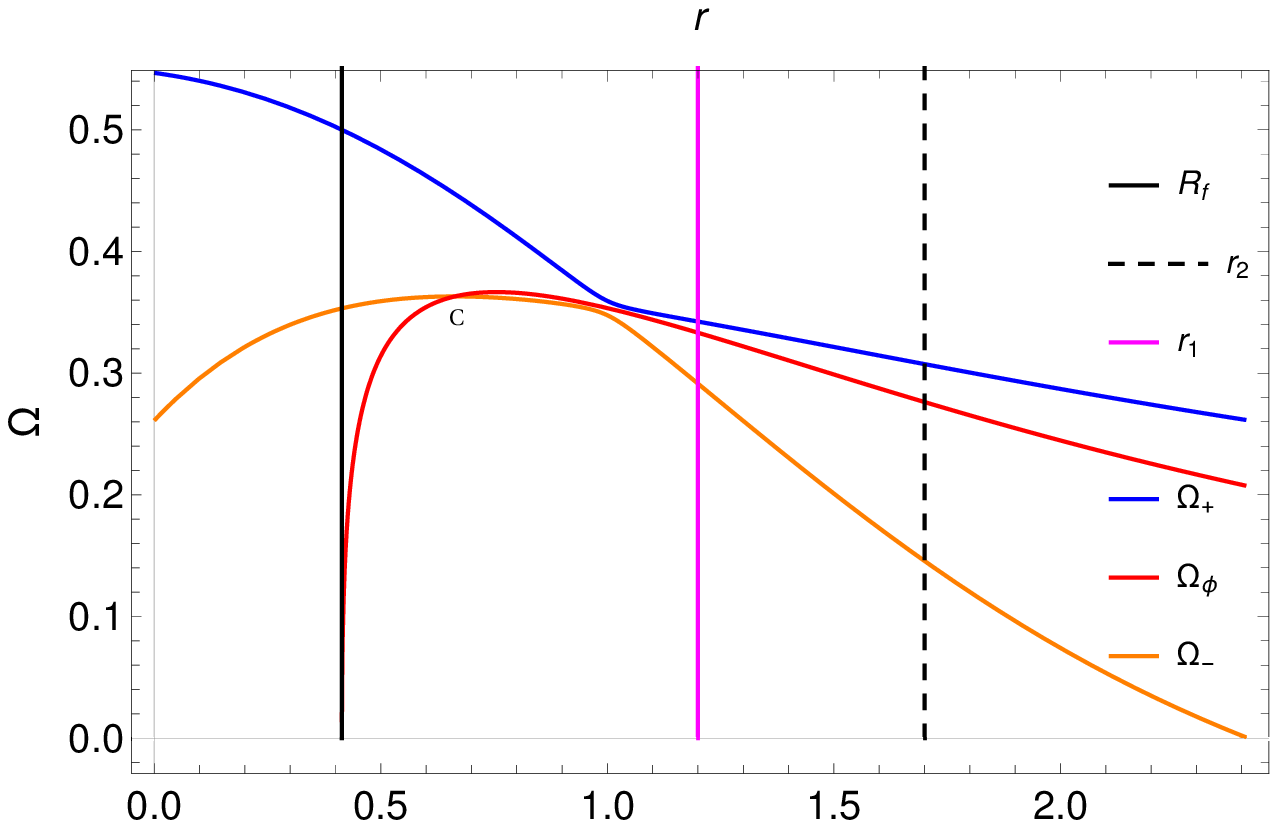}}
\hspace{0.05\textwidth}
\subfigure[KTN NS with $n_*=1$ and $a_*=1.45$ (corresponds to the orange curve of
figure \ref{ref})]{\includegraphics[width=2.8in,angle=0]{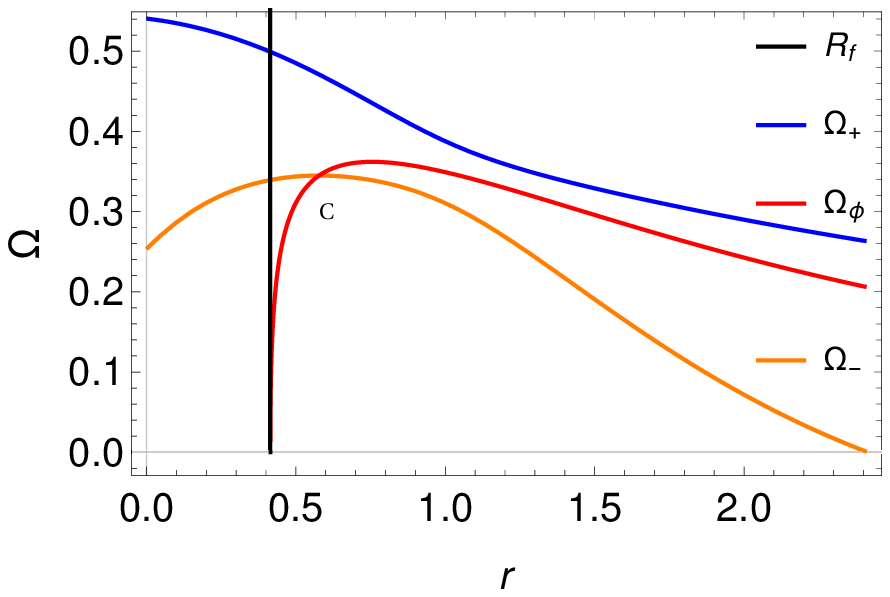}}
\hspace{0.05\textwidth}
\caption{\label{omega}Inside the ergoregion, a test particle can take only those $\O$
values which are in the following range : $\O_-< \O < \O_+$. $\O_+$ and $\O_-$ (in `$M^{-1}$') are
shown as solid blue and orange curves, respectively, and have been plotted specifically inside 
the ergoregion, in the equatorial plane ($\th=\pi/2$), as a function of $r$ (in `$M$').
For BHs, we can see from panels (a) and (c) that $\O_{\pm}$ meet at the horizon. 
For Kerr NS, we can see from panels (b) that $\O_{\pm}$ meet at the singularity ($r=0$) but 
they never meet in the case of KTN NSs (see panels (d) and (e)). The red solid curves
represent the Kepler frequency $\O_{\phi}$ which does not satisfy  
eq. (\ref{C})  after reaching the point `C', for the KTN NSs. We note, as no SCOs exist 
in the range $r_1 < r < r_2$ in panel (d), $\O_{\phi}$ is meaningless in this region.
See section \ref{av} for details.}
\end{figure*}

It is seen from panels (a) and (b) of figure \ref{omega} that this condition (eq. \ref{C}) is 
always satisfied in the cases of Kerr BH and Kerr NS. Panel (c) shows that it is
also satisfied in the case of a KTN BH but the condition mentioned in eq. (\ref{C})
cannot be satisfied in the case of a KTN NS (see panels (d) and (e)). This is because, it is clearly
seen from panels (d) and (e) that when the test particle reaches at the orbit of radius 
$r_{\rm C}$ (which corresponds to the point `C' of the plots drawn in panels (d) and (e)
of figure \ref{omega}), it coincides with
the $\O_-$ curve. Equating the expressions of $\O_{\phi}=\O_-$, one can numerically obtain
the value of $r_{\rm C}$. Remarkably, $r_{\rm C}$ coincides with the circular photon orbit (CPO)
{\footnote{See section \ref{cpo} for the detail discussions on the CPOs.}} or $r_c$ 
($r_{\rm C}\equiv r_c$) for every combination of $a_*$ and $n_*$ which represents a KTN NS. 
The reason behind that is, $\O_-$ is associated with the null vector $K_-=\p_t+\O_-\p_{\phi}$,
as was pointed out in eqs. (29--30) of \cite{ckp}. However, a test particle which
moves in a {\it timelike} circular orbit, unable to make its four-velocity $u$ become
{\it null} at $r_{\rm C}$. Therefore, the SCOs
which exist in the range : $r \leq r_{\rm C}$, cannot be accessible by a test particle
with its Kepler frequency. It would immediately fall into the KTN collapsed object 
and hence, the SCOs of $r \leq r_{\rm C}$ are not meaningful.

We may note here that $\O_{\pm}$ meet at the horizon and become a single-valued
function ($\O_{h}$), in case of a Kerr as well as a KTN BH. This frequency for
the KTN BH can be written as 
\begin{eqnarray}
 \O_h= \f{a}{2 (Mr_h + n^2)}.
\end{eqnarray}
It reduces to 
\begin{eqnarray}
 \O_h^{\rm Kerr}= \f{a}{2Mr_h^{\rm Kerr}}
\end{eqnarray}
(where $r_h^{\rm Kerr}/M=1+\sqrt{1-a_*^2}$ ) in the Kerr BH, which is well-known. 
For Kerr NS, $\O_{\pm}^{\rm Kerr}$ meet at the singularity ($r=0$) with 
$\O_{\pm}|_{r=0}=1/a$ \cite{ckp}, but for KTN NS $\O_{\pm}$ take two 
different values at $r=0$ :
\begin{eqnarray}
 \O_{\pm}|_{r=0} = \frac{2 a \pm \sqrt{a^2-n^2}}{n^2+3a^2}
 \label{opm0}
\end{eqnarray}
which is seen from panels (d) and (e) of figure \ref{omega}. We note that the condition 
$a^2-n^2 > 1$ holds for the KTN NS and therefore, we always obtain two values of 
$\O_{\pm}$ at $r=0$.
\\

{\bf Conclusion drawn from this discussion :} The above discussion tells us 
that a test particle cannot access the SCOs which exist in the region : 
$r \leq r_{\rm C}$ ($r_{\rm C}$ corresponds to the circular photon orbit)
for all four NS curves of figure \ref{ref}. Therefore, the `physical ISCO' 
should exist in the region : $r > r_{\rm C}$ and its exact location 
is obtained in the next section.

\subsubsection{Energy of the SCOs}\label{energy} 

\begin{figure}
   \begin{center}
\includegraphics[width=4in]{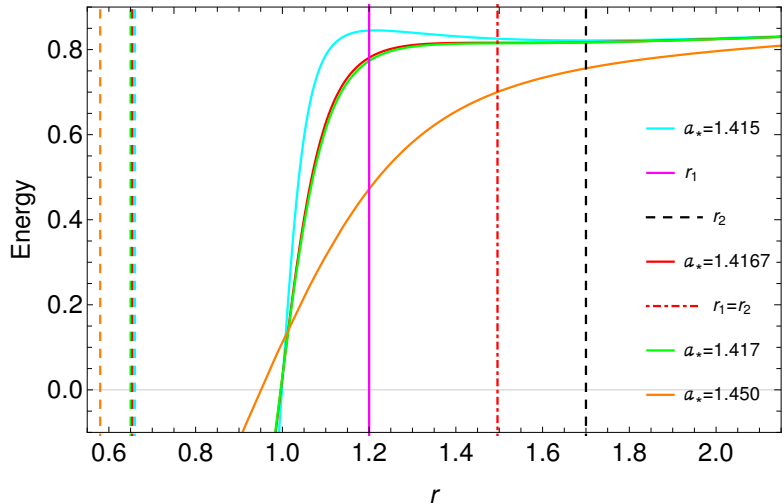}
     \caption{\label{en}Energy $E$ vs $r$ (in `$M$') for $n_*=1$ with  
     different values of $a_*$. The four dashed straight lines (cyan, red, green and orange) are the 
     corresponding $r_{\rm C}$ values ($r_{\rm C}/M=0.66, \,\, 0.654, \,\, 0.65, \,\, 0.58$) of the cyan, red,
     green and orange curves, respectively. Magenta and black dashed straight lines show the positions
     of $r_1$ and $r_2$ respectively, which arise only for the solid cyan curve. Note that 
    the cyan curve has actually no meaning in the region $r_1 < r < r_2$, as no SCOs exist 
    in this region. In case of the solid red curve, $r_1$ coincides with $r_2$, 
    i.e., $r_1=r_2$, which is represented by the dot-dashed red straight line.
    See sections \ref{energy} and \ref{sec:mb} for details.}
     \end{center}
\end{figure}

Increasing the energy $E$ with decreasing $r$
is one of the consequences of the absence of stable orbits \cite{jmn}.
One can check it in the region $r_1 < r < r_2$ for the cyan curve of figure \ref{en},
but, as $E$ decreases with decreasing of $r$ in the region $r < r_1$, we can 
conclude that SCOs exist in that region. Now, the cyan curve shows a sudden fall of $E$
for $r/M \sim 1.1$ and the test particle acquires $E=0$ at $r=r_0/M \approx 0.99$.
$E=0$ represents that the efficiency $(1-E)$ of accretion reaches $100\%$ at $r=r_0$ orbit,
i.e., all the mass-energy of the accreting gas is converted to 
radiation and returned to infinity (see second paragraph of section 3.2 of \cite{jmn}).
This implies a perfect engine which
converts mass into energy with $100\%$ efficiency \cite{jmn}.
Hence, the SCOs at $r/M < 0.99$, which have negative $E$ values,
are not meaningful. Therefore, in such a case, one can choose $r=r_0$ 
as the `physical ISCO'.

Figure \ref{en} is drawn for $n_*=1$ with different values of $a_*$. 
As is shown in this figure, $r_{\rm C}/M$ arise at $0.654, 0.65$ and $0.58$ for the red, green and orange 
curves, respectively. These are also far from the $E=0$ orbit ($r_0/M$) which occur at 
$0.9963, 0.9960$ and $0.95$ respectively for these three curves. 
Remarkably, one can see from figure \ref{ref}(a) and figure \ref{en} that there is no discontinuity in these three curves (including the solid red one), i.e., 
$E$ decreases with decreasing of $r$ in the whole range : $r_0 \leq r \leq \infty$
and becomes zero at $r=r_0$. 
Therefore, although one positive real root is obtained (by solving the so-called
ISCO equation) at $r_1=r_2=1.5M$ (red dot-dashed straight line) for the solid red curve, 
this may be ignored, and we can conclude that the `physical ISCO'
occurs at $r=r_0$ in all these three cases. Here, we should remember that these ISCOs do not carry the usual meaning which we
have discussed in section \ref{sec:isco}, as these new ISCO radii do not satisfy the so-called ISCO equation 
(eq. \ref{isco}). However, these new ISCO radii must satisfy the SCO condition mentioned in
section \ref{sec:isco}. One important thing is that there could exist many combinations of 
$a_*$ and $n_*$ (those are mainly fallen in the `red-coloured' region of figure \ref{noisco}),
for which the accretion efficiency reaches $100\%$ for the KTN NSs, whereas it is possible
only for those Kerr NSs whose $a_*$ values are in the following range: 
$1 < a_* \leq \sqrt{32/27}$. As an 
example, for $a_*=\sqrt{32/27}\approx 1.089$, accretion efficiency reaches $100\%$ at the ISCO : $r_{\rm ISCO}=2M/3$
\cite{gh} ($E$ becomes zero in this particular orbit, see \cite{ch,pug}).

\subsubsection{Importance of the marginally bound orbit}\label{sec:mb} 

\begin{figure}
\centering
\subfigure[Energy $E$ vs $r$ (in `$M$')]{\includegraphics[width=2.9in,angle=0]{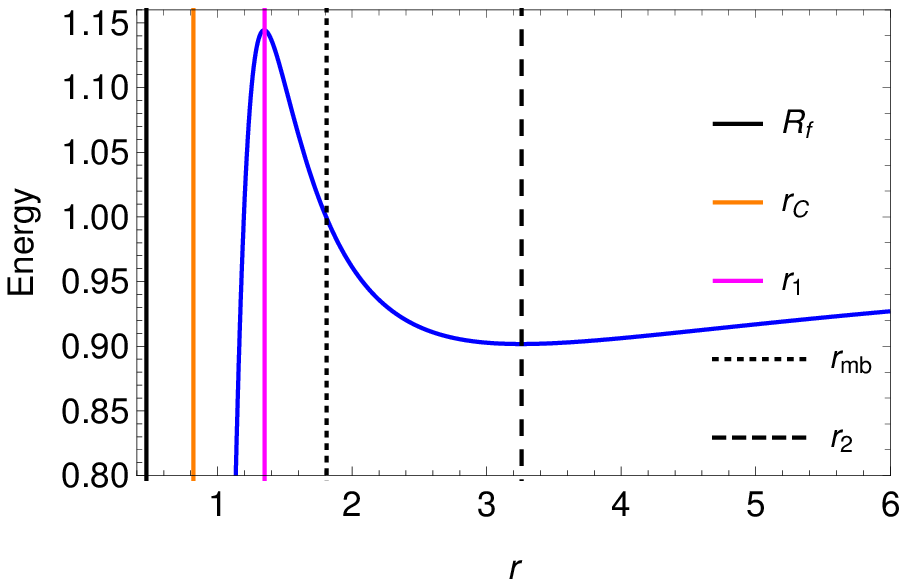}}
\hspace{0.05\textwidth}
\subfigure[$\O_r$ (in `$M^{-1}$') vs $r$ (in `$M$')]{\includegraphics[width=2.9in,angle=0]{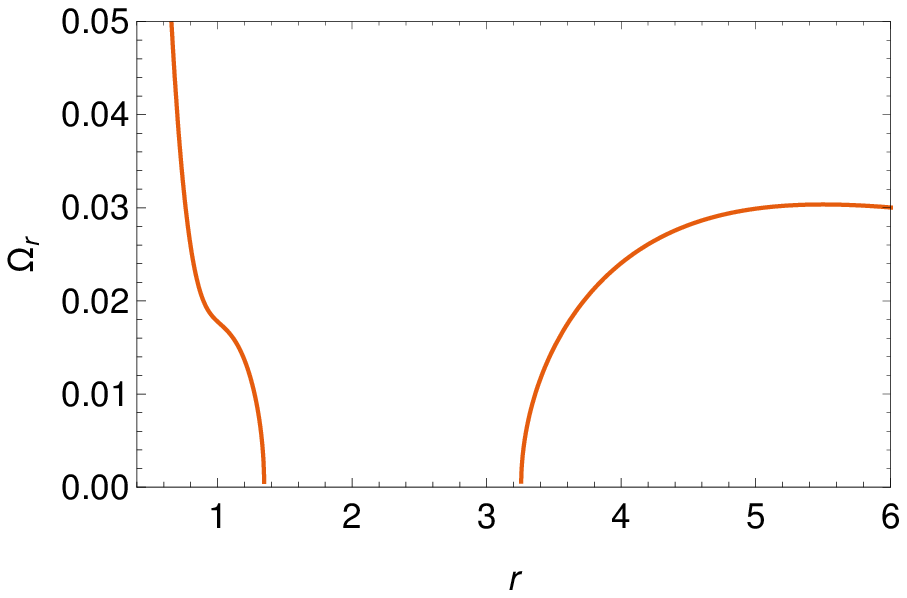}}
\subfigure[$V_{\rm eff}$ vs $r$ (in `$M$') for various values of $L$ (in `$M$') with 
$E=0.9$.]{\includegraphics[width=2.96in,angle=0]{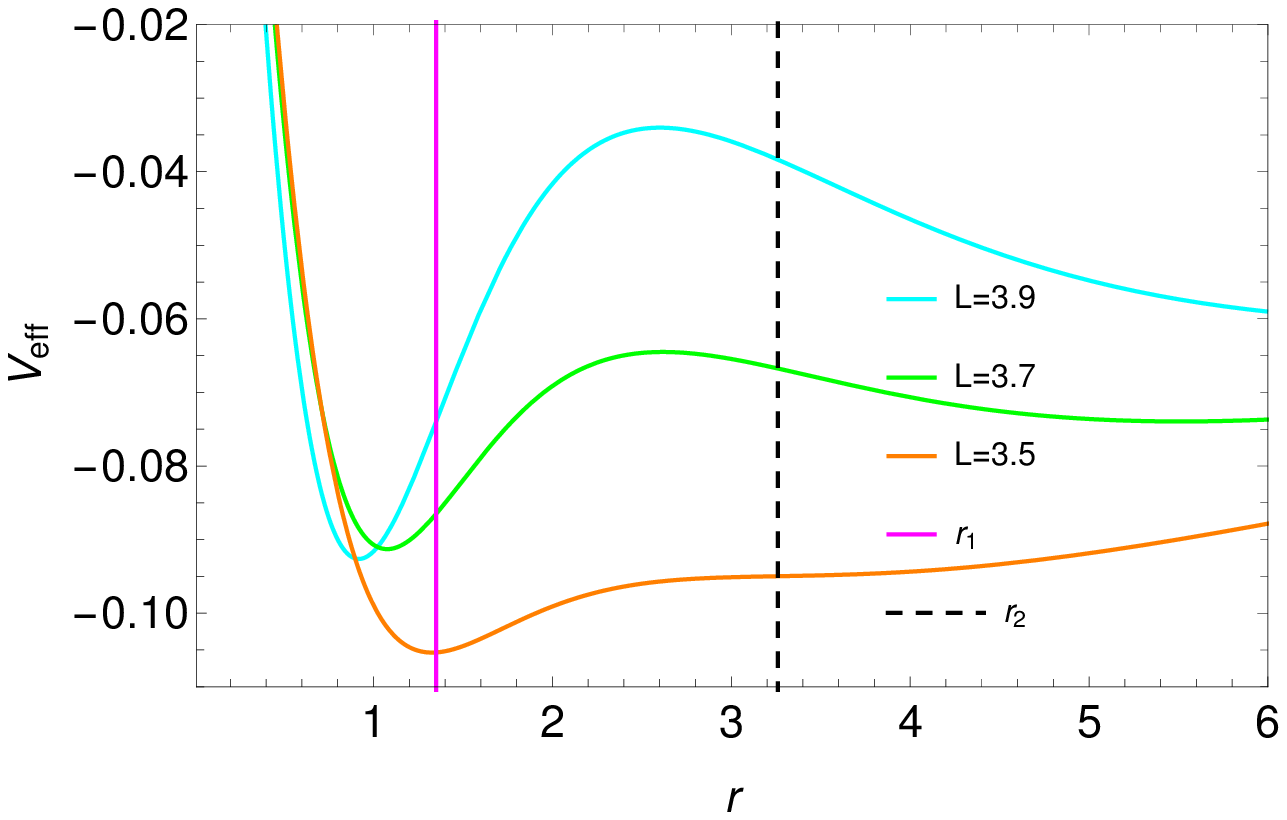}}
\hspace{0.05\textwidth}
\caption{\label{sc}Panel (a) shows the positions of various orbits for a KTN NS with $n_*=1.84$ and $a_*= 2.10$ 
: $R_f/M=0.47$,  $r_{\rm C}/M=0.82$, $r_1/M=1.35$, $r_2/M=3.26$ and $r_{\rm mb}/M=1.81$. Note that 
the blue energy curve has no meaning in the region $r_1 < r < r_2$, as no SCOs exist in 
this region, which is seen from panel (b). However, it is necessary to show the 
location of $r_{\rm mb}$. Panel (c) shows the behaviour of 
effective potential. See section \ref{sec:mb} for details.}
\end{figure}

Energy $E$ can increase or decrease with $r$ and depending on that one can find the location
of SCOs. It is seen from figure \ref{en} that the value of $E$ always remains less than $1$ in all
orbits for all curves. As we have discussed in section \ref{iscobh} that an orbit is called
marginally bound ($r_{\rm mb}$), if its energy becomes $1$. Now, setting $E=1$ in eq. (\ref{E}),
one can numerically solve the following expression
\begin{eqnarray}
(r^2+n^2)\left(r^3-3M r^2-3n^2r+M n^2+2a(m r)^{\f{1}{2}}\right)-
\left[r^{\f{1}{2}}(r^2-2M r-n^2)+a m^{\f{1}{2}}\right]^2=0 \nonumber
\\
\label{rmb}
\end{eqnarray}
to obtain the radius ($r_{\rm mb}$) of the marginally bound orbit. It is difficult 
to solve analytically for the KTN spacetime but can be solved for the Kerr BH,
which gives \cite{bpt} (see eq. 141 of Chapter 7 of \cite{ch} also)
\begin{eqnarray}
r_{\rm mb}^{\rm Kerr}=M \left(2-a_*+2\sqrt{1-a_*} \right)
\label{kmb}
\end{eqnarray}
for the direct orbits. One can check that the value of $r_{\rm mb}^{\rm Kerr}$ always
comes as less than the value of $r_{\rm ISCO}^{\rm Kerr}$ 
for a particular value of $a_*$ in case of the Kerr spacetime. For the extremal Kerr BH, $r_{\rm ISCO}^{\rm Kerr}$
and $r_{\rm mb}^{\rm Kerr}$ coalesce at the horizon : $r_h/M=1$.
That is why, we generally do not bother about $r_{\rm mb}^{\rm Kerr}$ for the 
realistic astrophysical problems in Kerr BH and deal only 
with the $r_{\rm ISCO}^{\rm Kerr}$.

Let us now consider figure \ref{sc} which is a special case in KTN spacetime. This figure 
corresponds to a KTN NS with $n_*=1.84$ and $a_*= 2.10$. One can safely say from panel (b)
of figure \ref{sc} that the SCOs do exist in the outer branch, i.e., $r_2/M \geq 3.26$. Panel (a)
also shows that $E$ of a test particle decreases with $r$ until
it reaches at $r_2/M=3.26$, whereas SCOs do not exist for the range 
$1.35 < r/M < 3.26$ as $E$ increase with $r$. This is also clear from panel (b). 
However, $r_{\rm mb}$ is located at $r_{\rm mb}/M=1.81$ and it palys an importnat
role in such a situation.
After reaching at $r=r_2$, the test particle should start to free-fall in principle
and reach at the $r=r_1=1.35M$ orbit, the feature of which is a bit similar to the 
cyan curve of figure \ref{en}. 
Incidentally, the energy of the $r_1$ orbit is greater than $1$ in this particular case, 
i.e., $E|_{r=r_1} > 1$,
which is seen from panel (a) of figure \ref{sc}. Moreover, one can also notice that the test particle 
faces the `unbound' ($E > 1$) orbits even in the
region : $r_1 < r < r_{\rm mb}$ before reaching at $r=r_1$. Therefore, as the test particle penetrates
to the unbound circular orbits ($E > 1$) at $r < r_{\rm mb}$, it must plunge directly into 
the collapsed object \cite{st} after crossing the $r=r_{\rm mb}$ orbit.
Therefore, the SCOs which occur at $r \leq r_1$ cannot be feasible at all and these orbits 
are also not meaningful for the accretion disk theory. In such a situation, $r_2$
should be considered as the `physical ISCO' and the inner branch is unfeasible.

\subsection{Special case}\label{sec:special}
\begin{figure}
\subfigure[$a_*=1.01$]{\includegraphics[width=3in,angle=0]{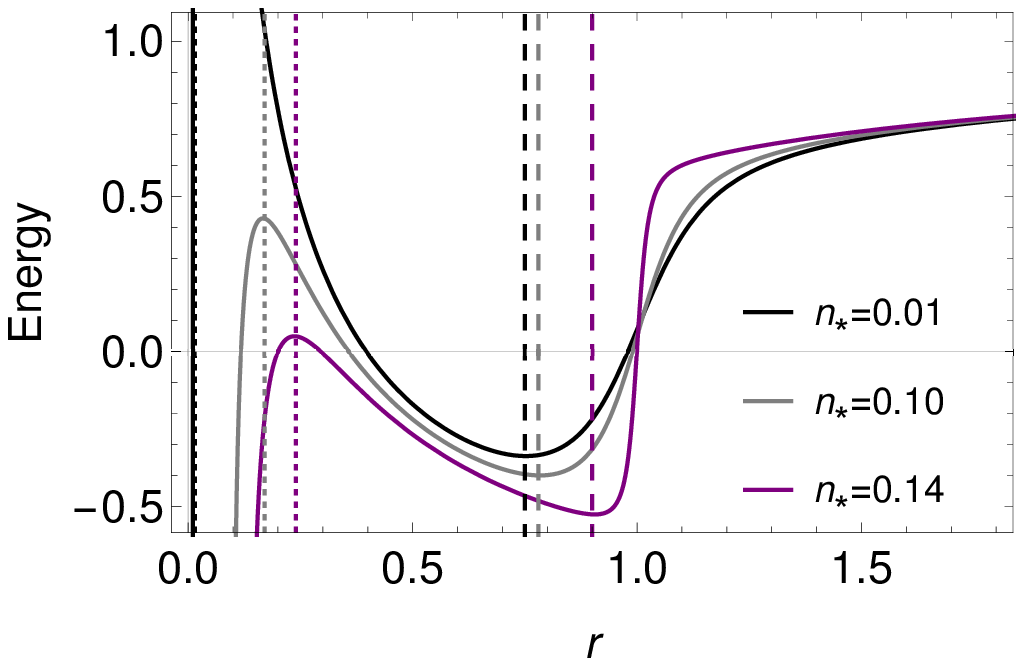}}
\hspace{0.05\textwidth}
\subfigure[$a_*=1.10$]{\includegraphics[width=3in,angle=0]{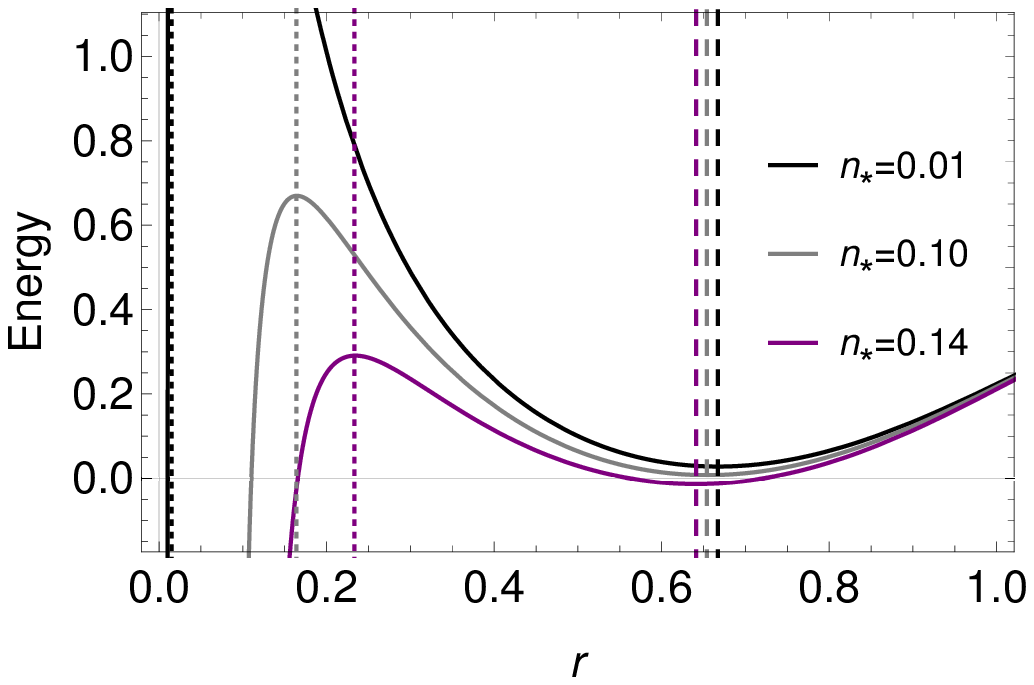}}
\subfigure[$a_*=1.30$]{\includegraphics[width=3in,angle=0]{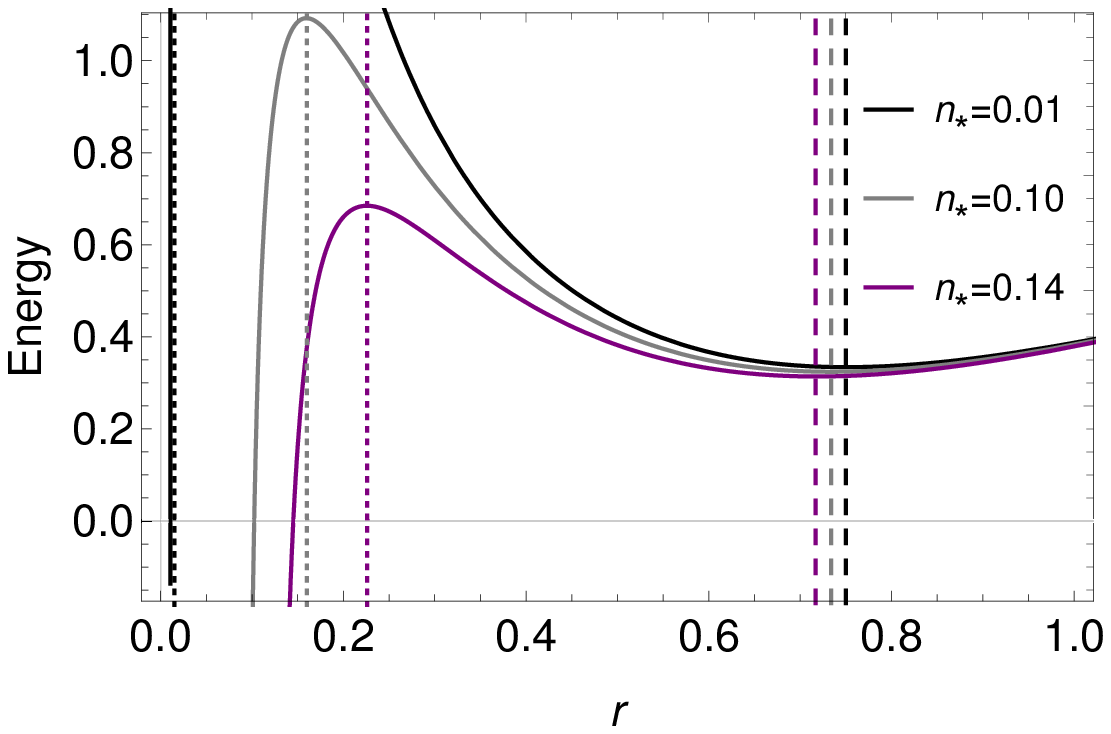}}
\subfigure[Zoomed version of panel (b) with range $0.50 < r/M < 0.85$
is shown for clarity. A new solid red curve is added for $n_*=0.1164$ to show that $r_{03}$
coincides with $r_{02}$ and $r_2$ (i.e., $r_{03}=r_{02}=r_2=0.651M$, the red 
dot-dashed straight line) in this particular case.]{\includegraphics[width=3in,angle=0]{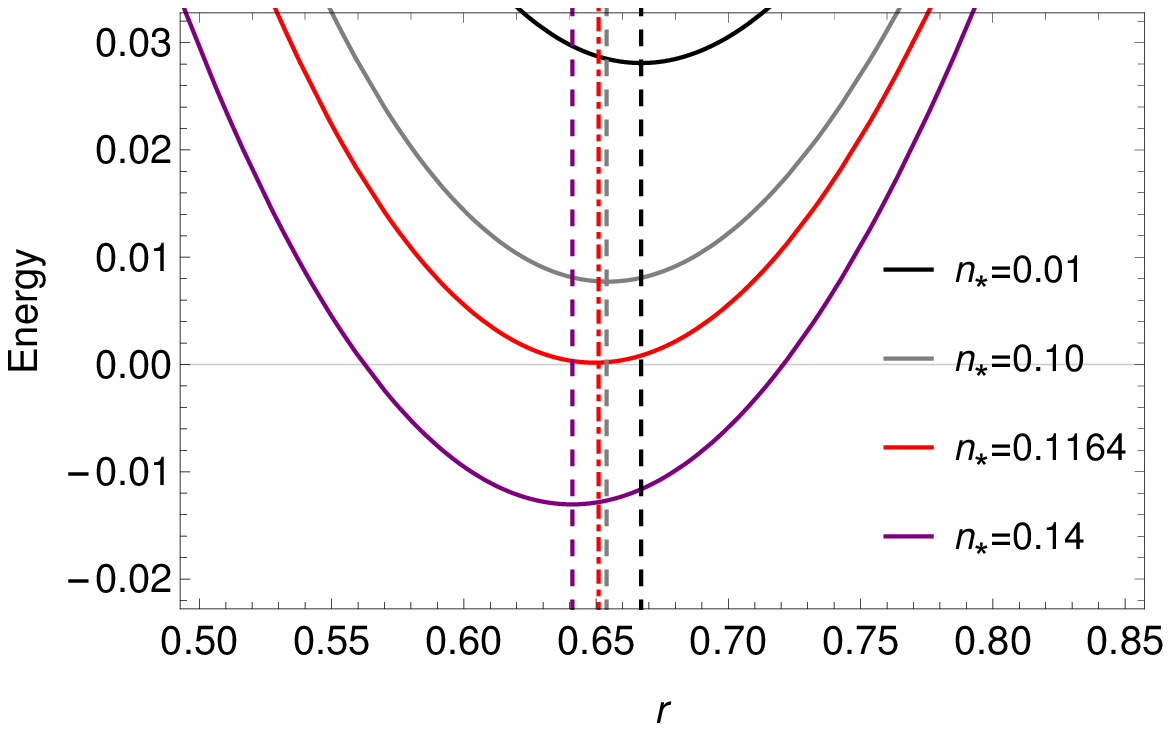}}
\hspace{0.05\textwidth}
\caption{\label{sac}Energy $E$ vs $r$ (in `$M$') for various values of $a_*$
with $n_*$ values close to zero. In each panel, the three dotted and dashed straight lines 
(black, gray, purple) are the corresponding $r_1$ and $r_2$ values of the black, gray and purple 
solid curves respectively, which represent the orbital energy. Although the $r_C$ values have not been 
shown in these plots, one can note that $r_C$ appears as $r_C < r_1$ in all these cases.
See section \ref{sec:special} for details.}
\end{figure}

One can see from figure \ref{en} that the value of $r_0$ comes as
smaller than the value of $r_2$, i.e, $r_0 < r_2$. In this section, we show that this is not always true. 
Let us consider figure \ref{sac}. Panels (a)-(c) of the same are drawn for three different values of 
$a_*=1.01, 1.10$ and $1.30$ with the $n_*$ values are almost close to zero, i.e., $n_*=0.01, 0.10$ 
and $0.14$. In each panel, the three dotted and dashed straight lines 
(black, gray, purple) are the corresponding $r_1$ and $r_2$ values of the black, gray and purple 
solid curves respectively, which represent energy of the circular orbits. One can see that
feature of the three
solid energy curves of panel (a) of figure \ref{sac} is completely different from the feature
of the energy curves of figures \ref{en} and \ref{sc}(a). In figure \ref{sac}, each energy curve not only crosses $E=0$ axis for three
different values of $r$ (say, $r_{01}, r_{02}$ and $r_{03}$, such that 
$r_{01} < r_{02} < r_{03}$), but also one of them ($r_{03}$) appears before $r_2$, i.e., $r_2 < r_{03}$.  
Now, consider a test particle/accreting matter approaches to a KTN NS of $a_*=1.01$ and
$0 < n_* \lesssim 0.142$ from infinity (in principle). It is needless to say that it encounters $r=r_{03}$ 
orbit first instead of $r=r_2$ orbit unlike the cyan curve of figure \ref{en}. As all the mass-energy of the accreting
gas is converted to radiation and returned to infinity \cite{jmn} from this $r=r_{03}$ orbit, no orbits with
$r < r_{03}$ are feasible in these cases. Therefore, $r=r_{03}$ should be considered as the 
`physical ISCO' for all these three cases of panel (a) of figure \ref{sac}, i.e., 
$r_{03}/M=0.97, 0.98$ and 
$0.99$ for black, gray and purple solid curves respectively. Here, we should note that 
$n_* \geq 0.142$ with $a_*=1.01$ represent KTN BHs and one has to follow the discussion of 
section \ref{iscobh} to determine the location of ISCO for those cases.

Now, if we increase the value of $a_*$ from $1.01$ to $1.10$ (without changing the values of
$n_*$), the special feature of the energy curves of figure \ref{sac}(a) starts to disappear 
(see panel (b)) depending on the values of $n_*$ and it completely disappears for higher values of $a_*$
which can be seen from panel (c). We can see from panel (b) of figure \ref{sac}
that although the solid purple curve crosses the $E=0$ axis thrice (similar to the 
solid curves of panel (a)), the gray and black curves cross the same only once
(similar to the solid curves of figure \ref{en}). This is also clear from panel (d) of figure \ref{sac},
which is the zoomed version of panel (b). Therefore, if a test particle/accreting matter
approaches to a KTN NS of $a_*=1.10$ and $n_*=0.14$ (purple curve) from infinity (in principle), it first encounters 
$r_{03}/M=0.72$ (see, $r_{03} > r_2$) orbit which should be the `physical ISCO' as is described in the previous
paragraph. A new solid red curve is added in panel (d) for the KTN NS of $n_*=0.1164$ and 
$a_*=1.10$ to show that $r_{03}$ coincides with $r_{02}$ and $r_2$  at $r/M=0.651$ 
(i.e., $r_{03}=r_{02}=r_2=0.651M$, the red dot-dashed straight line)
which should be considered as the `physical ISCO' in this case. However, the characteristic of solid gray
curve of panel (b) and the solid purple curve of panel (c) of figure \ref{sac} is similar to  
the solid cyan curve of figure \ref{en}. Therefore, one should consider the $E=0$ orbit
(i.e., $r_0/M=0.11$ and $r_0/M=0.14$ respectively) as the `physical ISCO' in these two cases.
The characteristic of the gray curve of panel (c) resembles to the solid blue curve of figure
\ref{sc}(a) and one should consider $r_2/M=0.73$ as the `physical ISCO' in this case.
One can see that the two solid black curves of panels (b) and (c) of figure \ref{sac} continue
to very close to $r=0$ and $r_1$ (i.e., dotted black straight lines) occurs just outside of it. 
The characteristic of these two curves are also similar to the solid blue curve of figure
\ref{sc}(a) and $r=r_2$ is considered as the `physical ISCO', as discussed in section \ref{sec:mb}.

\section{Circular photon orbits}\label{cpo}
In this section, we study the location of the circular photon orbit in KTN spacetime,
for completeness. The expressions of $L$ and $E$ (eqs. \ref{E}-\ref{L}) imply
that the circular orbits exist down in 
a particular orbit of radius $r=r_{\rm c}$ which is the solution of the following equation
\begin{eqnarray}
 r_{\rm c}^3-3M r_{\rm c}^2-3n^2 r_{\rm c}+M n^2+2a(m_{\rm c} r_{\rm c})^{\f{1}{2}}=0
 \label{po}
\end{eqnarray}
where $m_{\rm c}=M~(r_{\rm c}^2-n^2)+2~n^2r_{\rm c}$. Specifically, $E$ becomes
infinity at $r=r_{\rm c}$ and hence the solution of eq. (\ref{po}) gives the
radius of CPO $r=r_{\rm c}$. Eq. (\ref{po}) is difficult
to solve analytically but we can easily solve it numerically.
\\

\subsection{CPOs in KTN black holes}\label{cpobh}
One can obtain only one positive real root of eq. (\ref{po}) outside the horizon in 
case of a non-extremal KTN BH ($a_* < \sqrt{1+n_*^2}$), which
implies that only one CPO occurs for each corresponding combinations of 
$a_*$ and $n_*$ and the CPO must exists outside the event horizon.

In case of an extremal KTN BH $\left(a_*= \sqrt{1+n_*^2} \right)$
horizon occurs at $r_h=M$ and the expression (eq. \ref{po}) of CPO with radius 
$r_{\rm cx}$ reduces to
\begin{eqnarray}
 r_{\rm cx}^3-3M r_{\rm cx}^2-3n^2 r_{\rm cx}+M n^2+2\sqrt{m_{\rm cx} r_{\rm cx}(M^2+n^2)}=0 
 \label{poex}
\end{eqnarray}
where, $m_{\rm cx}=M~(r_{\rm cx}^2-n^2)+2~n^2r_{\rm cx}$.
The analytical solutions of the above equation are
\begin{eqnarray}\nonumber
 r_{\rm cx}&=& M, \,\,  M \left(1 \pm \sqrt{1+n_*^2} + \sqrt{2} \sqrt{1+n_*^2 \pm \sqrt{1+n_*^2}} \right) ,
 \\ 
  && M \left(1 \pm \sqrt{1+n_*^2} - \sqrt{2} \sqrt{1+n_*^2 \pm 1\sqrt{1+n_*^2}} \right)
\end{eqnarray}
of which 
\begin{eqnarray}\nonumber
&& r_{\rm cx}^1 =M 
\end{eqnarray}
and,
\begin{eqnarray}
r_{\rm cx}^{\mp}=M \left( 1 \mp \sqrt{1+n_*^2} 
 + \sqrt{2} \sqrt{1+n_*^2 \mp \sqrt{1+n_*^2}} \right) 
\end{eqnarray}
are positive and hence these are acceptable. One of these occurs at $r_{\rm cx}^1=M$ 
due to the prograde rotation whereas $r_{\rm cx}^+$ occurs due to the retrograde rotation.
A short calculation reveals that the $r_{\rm cx}^-$ occurs inside the event 
horizon ($r_{\rm cx}^- < r_h=M$) for $n_* < \sqrt{3}$ (i.e., $a_* < 2$) but it occurs outside of 
the event horizon for $n_* > \sqrt{3}$ (i.e., $a_* > 2$). 
This means that two CPOs could exist simultaneously for $n_* > \sqrt{3}$, whereas
one can neglect $r_{\rm cx}^-$ for $n_* < \sqrt{3}$, as it occurs inside the event horizon.
In this special case ($n_* > \sqrt{3}$), one CPO coincides with the event horizon and the 
other one exists outside of $r_h$. For $n_* = \sqrt{3}$ (i.e., $a_* = 2$), 
$r_{\rm cx}^1=r_{\rm cx}^-=M$,
hence, only one CPO occurs in this case. One can check that $r_{\rm cx}^-$ becomes $0$ 
for $n_*=0$. That is why, we cannot obtain two CPOs for the extremal Kerr 
BH.{\footnote{For $n_*=0$, $r_{\rm cx}^+$ reduces to $4M$ which indicates that the 
CPO occurs at $4M$ for the retrograde rotation in the extremal Kerr BH case.}} 
\\

\subsection{CPOs in KTN naked singularities and comparison with Kerr spacetime}

For each corresponding combinations of $a_*$ and $n_*$, we always obtain only
one positive real root of eq. (\ref{po}) in case of a KTN NS,
and it occurs at $0 < r_c < M$.

It could be useful to mention here the location of corotating CPOs in Kerr spacetime
for comparison.
For a Kerr BH ($0 < a_* \leq 1$), one can calculate the
radius of the corotating CPO as $3M > r_{\rm c}^{\rm Kerr} \geq M$ using the following equation \cite{ch, sti} :
\begin{eqnarray}
 r_{\rm c}^{\rm Kerr}=2M \left[1+\cos \left\{\f{2}{3}\cos^{-1}\left(-a_* \right)\right\}\right].
 \label{pok}
\end{eqnarray}
The corotating CPO does not occur in case of a Kerr NS (CPO is formally 
located at the ring singularity) \cite{chastu} but remarkably it occurs for a KTN NS
at $0 < r_c < M$.
This could be an interesting distinguishable feature between a Kerr NS and a KTN NS. 
We can safely say that the non-zero value of $n_*$ is basically 
responsible for this. Because, setting $a \rightarrow 0$ in eq. (\ref{po}), 
one can check that the CPO can always occur at \cite{cc2, jp}
 \begin{eqnarray}\nonumber
r_{c}^{\rm TN} \rightarrow M\left[1+2(1+n_*^2)^{1/2}\cos\left\{\f{1}{3}\tan^{-1}\left(n_*\right)\right\}\right]
\\
\end{eqnarray}
(TN stands for Taub-NUT spacetime) for all values of $n_*$, in principle. 

It is well-known that the ISCO occurs at
$6M > r_{\rm ISCO}^{\rm Kerr} \geq M$. Hence, the CPO and ISCO coalesce on the 
horizon ($r_h=M$) only in the case of extremal Kerr BH $(a_* =1)$ \cite{ch}.
Otherwise, the ISCO always occurs outside 
the CPO for every value of $a_*$, i.e., $ r_{\rm ISCO}^{\rm Kerr} > r_{\rm c}^{\rm Kerr}$ for
$0 \leq a_* < 1$.

\section{Conclusion and Discussion}\label{dis}
Here we report several new and interesting features of circular orbits in the KTN 
spacetime. We have derived the radii of CPOs as well as SCOs and discussed
their implications for the accreting KTN BHs and KTN NSs.
We have mainly shown that the location of ISCO for the non-extremal KTN BH can easily be determined 
by solving eq. (\ref{isco}), but one cannot determine the same only by solving
the so-called ISCO equation (eq. \ref{isco}) for the extremal KTN BH and KTN NS.
Some SCOs are unfeasible due to the various reasons as is discussed in sections
\ref{fr}--\ref{sec:mb}. The intriguing behaviour of the Kepler frequency and
the restriction on angular velocity inside the KTN ergoregion are the main
reasons for this. Above all, the orbital energy plays the most vital role to determine
the location of `physical ISCO'.
We have calculated a few numerical values of the `physical ISCO' radii in sections
\ref{energy}--\ref{sec:mb} and section \ref{sec:special}, but one can easily 
calculate the numerical value of the `physical ISCO' for any specific combination of $a_*$
and $n_*$, following the same procedure which we have pursued in sections \ref{fr}--\ref{sec:mb}.
In reality, the infinite combinations of $a_*$ and $n_*$ is possible (in principle),
and one can catagorize it into four different types.

(i) Solving the so-called ISCO equation, two positive real roots are obtained for an
extremal KTN BH. One occurs at $r=r_1=M$ (on the event horizon) for all values of $n_*$ and another
occurs at $r=r_2$ which depends on the values of $n_*$. We have shown that the $r_1$ is unfeasible 
for the extremal KTN BHs with $n_* > 0.577$, and therefore $r=r_2$ should be regarded
as the `physical ISCO' in these cases, whereas $r=r_1=M$ should be the `physical ISCO'
for the extremal KTN BHs with $0 < n_* \lesssim 0.577$.

(ii) Similar to Case (i), one can also obtain two positive real roots ($r_1$ and $r_2$) of the 
so-called ISCO equation as the solution of one class of KTN NSs. From the plot of radial 
epicyclic frequency, one can clearly see that two branches of SCOs
appear in the range of $r$, i.e.,  $r : R_f < r \leq r_1$ (inner branch) and 
$r : r_2 \leq r < \infty$ (outer branch).  
Although, all SCOs in the outer branch are feasible, this is not true for all SCOs of
the inner branch. If the energy of any orbit is greater than $1$ (i.e., $E > 1$) in the region:
$r_1 < r < r_2$, the particle/matter must plunge directly into the collapsed object
after crossing the $E=1$ orbit (i.e. $r=r_{\rm mb}$) as it is marginally bound. 
In such a situation, we have to identify $r_2$ as the `physical ISCO', 
as the whole inner branch is unfeasible in this particular case.
If the particle/matter does not face $E > 1$ orbits in the region $r_1 < r < r_2$,
and the energy of the $r_1$ orbit is less than or equal to $1$
(i.e., $E \leq 1$), the test particle or the matter makes its motion stable further 
in the SCOs of inner branch. However, as the energy of SCOs decreases with decreasing 
the value of $r$, the matter faces $E=0$ (i.e., $r=r_0$) orbit.
In this case, $r=r_0$ should be considered as the `physical ISCO'. 

Now, if the two positive real roots coincide ($r_1=r_2$), $r=r_0$ should be the `physical ISCO'
as the energy of SCOs decreases with decreasing the value of $r$ and it reaches $E=0$ at $r=r_0$.

(iii) In Case (ii), the test particle/matter first encounters the $r_2$ orbit before facing the 
$r_0$ orbit,
i.e., $r_2 > r_0$. In some cases, it is also possible to encounter the $r_0$ orbit first
instead of $r_2$. In such a situation, the $r=r_0$ orbit should be considered as the 
`physical ISCO'. It is needless to say here that if $r_0$ and $r_2$ coincide ($r_0=r_2$),
the same has to be considered as the `physical ISCO'.

(iv) The most interesting case is when one does not obtain any positive real root of the 
so-called ISCO equation, as the solution of one particular class of KTN NSs. In this case,
the test particle does not face $E > 1$ orbits but it must face the $E=0$ orbit, 
as the energy decreases with decreasing $r$ in the SCOs. We have already shown that 
in such a situation, accretion efficiency reaches to $100\%$ at the $r=r_0$ orbit, 
and hence, we choose $r=r_0$ as the `physical ISCO' in this case.

One should note here that like $a_*$ the value of $n_*$ can be different 
for different objects and it can even be very close to zero for 
some objects \cite{cbgm}. Finally, we emphasize that the SCOs (in the equatorial plane) are more relevant when the accretion 
disk around a collapsed object is geometrically thin and Keplerian. 
Such a disk is expected to have a multicolour blackbody spectrum, and from such an observed spectral 
component, this type of disk can be inferred \cite{RemillardMcClintock2006}. For such a disk, SCOs, including ISCO, can be important 
to interpret the observed quasi-periodic oscillations (QPOs) \cite{bs14, cb17, bcb18}. The energy and shape of the broad 
relativistic iron line \cite{Miller2007} observed from accreting collapsed objects are also expected to depend on the
disk inner edge radius, and hence on SCOs and ISCO.
\\

{\bf Acknowledgements :} 
We thank S. Ramaswamy and P. Majumdar for useful discussions on this topic.
We thank the anonymous referee for the constructive comments and valuable 
suggestions that helped to improve the manuscript.
C. C. gratefully acknowledges support from the National Natural Science Foundation 
of China (NSFC), Grant No.: 11750110410 and the China Postdoctoral Science Foundation,
Grant No.: 2018M630023.

\end{document}